\RequirePackage{fix-cm}
\documentclass[a4paper,11pt]{article}
\pdfoutput=1 

\usepackage{jinstpub} 
\usepackage{lineno}

\title{Analytical model for the photomultiplier single photoelectron response including the electron back-scattering contribution.}

\author[1]{E.~Angelino,}
\author[2,a]{V.~Beligotti,}
\author[2]{L.~Bellagamba,}
\author[b,2]{E.~Bonali,}
\author[2]{G.~Bruni,}
\author[2]{P.~Di Gangi,}
\author[b,2]{G.~Lucchetti,}
\author[b,2]{A.~Mancuso,}
\author[b,2]{V.~Mazza,}
\author[b,2]{G.~Sartorelli,}
\author[2]{M.~Selvi,}
\author[2]{F.~Semeria,}
\author[1]{A.~Razeto,}
\author[3]{S.~Vecchi,}
\author[c,3]{G.~Zavattini}
\affiliation[1]{INFN, Laboratori Nazionali del Gran Sasso, Assergi 67100, Italy}
\affiliation[2]{INFN, Sezione di Bologna, Bologna 40126, Italy}
\affiliation[3]{INFN, Sezione di Ferrara, Ferrara 44122, Italy}
\affiliation[a]{Gran Sasso Science Institute, L'Aquila 67100, Italy}
\affiliation[b]{Dipartimento di Fisica e Astronomia, Universit\`a degli Studi di Bologna, Bologna 40126, Italy}
\affiliation[c]{Dipartimento di Fisica e Scienze della Terra, Universit\`a degli Studi di Ferrara, Ferrara 44120, Italy}

\emailAdd{guido.zavattini@unife.it}

\abstract{
Many models exist to describe the single photoelectron response of photomultipliers. Generally to describe the spectral region between the fully amplified primary photoelectron peak and the electronics pedestal an \emph{ad hoc} function is used (often an exponentially modified gaussian) without a clear physical justification.
In this paper, following the physical description of back-scattered primary photoelectrons at the first dynode described in the "The Photomultiplier Handbook" by A.~G. Wright published by Oxford University Press, we derive an analytical function describing these partially amplified primary photoelectrons. This function depends only on intrinsic parameters of the photomultiplier such as the gain at the first dynode and the intrinsic resolution of the dynode chain following the first. Furthermore, analytical descriptions of the fully amplified peak and very low charge signals are derived.
The model has been successfully validated with data from two different photomultipliers acquired with a low-noise amplifier.
}
\arxivnumber{2604.02826}
\begin{document}

\maketitle

\section{Introduction}
\label{sec:intro}
Although new single photon light sensors have been developed during the last 30 years, photomultipliers  remain a robust and excellent choice in many applications in which very low light levels need to be detected over large surface areas. 

In general, for any single-photon light detector the single photoelectron (SPE) response is a fundamental characteristic which needs to be modelled in order to determine the effective acceptance, gain, resolution \emph{etc.} of the light detector.  

Extensive literature can be found \cite{Wright,Bellamy1993,Dossi2000,DEAP2019,IceCube2020,Kaloisis2024} (and others within) describing the SPE response but none are fully satisfactory as there is always some {\it ad hoc} assumption made to justify the measured SPE responses especially for signals between the pedestal peak and the single photoelectron peak. 
Generally, the SPE response results as the combination of two components: a clear peak related to the photoelectrons emitted by the photocathode which are completely amplified by the dynode chain, and a significant low-energy contribution due to undersized signals.
The peak is well described by a continuous, scaled, Poisson distribution due to the first dynode gain fluctuations and Gaussian broadening of the remaining dynodes. 
The signals below the fully amplified peak are  often described by an exponentially modified Gaussian or other similar function.

In this paper we propose the implementation of a model of energy loss of electrons in the first dynode presented in ``The Photomultiplier Handbook'' by A.~G. Wright \cite{Wright} in which the energy distribution of the back-scattered electrons is described. This process is also described in literature especially in the realm of electron microscopy \cite{DEKKER1958,Reimer1998}. Indeed, a significant fraction of primary electrons reaching the first dynode are back-scattered (see for example \cite{Wright,Assa'd}) depositing only a fraction of their energy in the dynode available for generating secondary electrons.

\section{SPE response model}\label{sec:section2}
In this section we present the details of the SPE model proposed to describe the response of an $N$-dynode photomultiplier. 

The electron multiplication process is separated into two stages:
%
the first dynode and the subsequent $N-1$ dynodes. 
The amplification at the first dynode determines the overall shape of the SPE spectrum. Since the gain of this stage drives the quality of the amplification process, it is typically set as high as possible. 
The amplification in the remaining $N-1$ dynodes contribute to a slight smearing of the spectrum generated at the first dynode, which is approximately Gaussian.
In addition, the $N-1$ dynode chain is responsible for very-low charge signals. 

In the following, the responses of the first dynode and of the subsequent dynode chain are discussed separately. Interestingly, we find that the overall SPE response of a photomultiplier can be accurately described using only a small number of parameters.



\subsection{SPE response at the first dynode}
The SPE response at the first dynode has two dominant contributions: fully amplified (FA) and partially amplified (PA) primary electrons.
\subsubsection{Fully amplified photoelectrons}
The voltage between the photocathode and first dynode determines the gain at the first dynode, $G_1$. 
Depending on the dynode material, gains of up to $G_1 \sim 30$ can be achieved. These gains refer to the condition in which the primary photoelectron deposits all of its energy in the dynode (no back-scattering) generating on average $G_1$ secondary electrons. 
In this situation the emission of $n$ secondary electrons follows Poisson statistics with mean and variance $G_1$:
\begin{align}
\label{eq:Poisson}
P_{\rm FA}(n,G_1) = \frac{e^{-G_1}G_1^n}{n!}.
\end{align}
\subsubsection{Partially amplified photoelectrons at the first dynode}
Not all photoelectrons generated at the photocathode are fully amplified at the first dynode with gain $G_1$. Several effects compete to this: inelastic and elastic back-scattering, bad collection of the primary photoelectrons from the photocathode, non-uniformity of the dynode gain $G_1$, \emph{etc}. Among these, the dominating contribution to partially amplified photoelectrons is inelastic back-scattering at the first dynode, with a fraction  $\eta\gtrsim 30\%$~\cite{Wright,Assa'd}. In this process the photoelectron's energy is only partially deposited in the first dynode resulting in a reduced number of secondary electrons being generated. If the back-scattered electron is lost, partially amplified signals are generated and the effective gain at the first dynode for this component is reduced: $1\le G_1^{\rm (eff)}<G_1$.\footnote{Note that the effective gain can only be $\ge 1$ given the discrete nature of electrons.} 
If the electric field recovers the back-scattered electron, the whole electron's energy is deposited on the first dynode, resulting in a fully amplified signal. 
This category of events can be distinguished thanks its peculiar time-structure, which can be resolved by acquiring the signal waveforms with a dedicated timing apparatus~\cite{LUBSANDORZHIEV2000,LUBSANDORZHIEV2006,KAETHER2012}.

Here we discuss the case in which the back-scattered electron is lost. The model implemented in the following sections is derived from Ref.~\cite{Wright} and in particular Chapter 5, Section 5.5 which we repeat here for completeness. 

Figure 5.8 of Ref.~\cite{Wright} shows the experimentally determined general trend of the energy spectrum of the electrons leaving the first dynode when hit by primary electrons accelerated to a kinetic energy of $E_p$~\cite{DEKKER1958,Wright}. Three regions are indicated:
\begin{itemize} 
\item {\bf Region A} shows the energy spectrum of secondary electrons \emph{emitted} by the first dynode subsequent to energy deposited by a primary photoelectron, either fully or partially amplified. 
\item {\bf Region B} shows the uniform energy spectrum\footnote{As Wright mentions (Ref.~\cite{Wright}, Sec.5.5.1), two comments must be made: firstly the Region B spectrum in reality shows a slightly dished distribution, and secondly the experimental spectrum for highly active alkali  materials, of interest in this paper, has not been measured \cite{Wright}. Nonetheless, here a uniform spectrum is assumed, considering that this approximation is not critical in the following model.} 
of (inelastic) back-scattered electrons. 
The back scattered electron therefore carries away an energy $E'_p$ smaller than $E_p$. 
The deposited energy $E_p-E'_p$ will generate a certain number of secondary electrons contributing to Region A. 
The fraction of these events, $\eta$, for high $Z$ metals can be $\eta\gtrsim 0.3$, especially for values of $E_p \gtrsim 600$~eV as is often the case in single-photon counting photomultipliers. 
\item {\bf Region C} shows elastically back-scattered primary electrons which do not generate secondary electrons and are either lost or redirected to the first dynode to start the amplification process over again (either fully or partially amplified). These events represent only a few percent of the cases~\cite{Wright}.
\end{itemize}

What follows is dedicated to a model describing those events in Region B which generate a reduced number of secondary electrons.

In general the mean number of secondary electrons generated following an energy deposit $E$ in the first dynode is $E/\epsilon$, where $\epsilon$ is the energy necessary to generate one secondary electron. Given an incident primary electron with kinetic energy $E_p$, in a fraction $\eta$ of cases, the primary photoelectron carries away an energy $E'_p = \kappa E_p$ due to inelastic scattering and, therefore, deposits an energy $E_p - E'_p = \left(1-\kappa\right) E_p$ in the dynode for secondary emission. In this case $\kappa$ is uniformly distributed between 0 and 1 given that the spectrum in Region B is uniform. The number $n$ of secondary electrons follows a Poisson distribution with mean $  \left(1-\kappa\right) E_p/ \epsilon = \left(1-\kappa\right) G_1 $:
\begin{align}
P\left(n,G_1\left(1-\kappa\right)\right) = e^{-G_1\left(1-\kappa\right)}\left(G_1\left(1-\kappa\right)\right)^n/n! \, .
\end{align}
Integrating $P\left(n,G_1\left(1-\kappa\right)\right)$ over all values of $\kappa$ one finds the following probabilities for various values of $n$: 
\begin{equation}
    {\cal P}\left(n,G_1\right) =\frac{1}{G_1}\left[1-e^{-G_1}\sum_{k=0}^n\frac{{G_1}^k}{k!}\right].
\label{eq:cassetta}
\end{equation}
These secondary electrons contribute to the partially amplified signals in a SPE spectrum. To be detected, though, only values of ${\cal P} (n,G_1)$ for $n>0$ must be considered and Eq. \eqref{eq:cassetta} must then be re-normalized by a factor $F = 1-(1-e^{-G_1})/G_1$. The probability density function for detectable secondary electrons is therefore
\begin{align}
\label{eq:cassetta_norm}
    P_{\rm PA} (n,G_1) = {\cal P} (n,G_1)/F  \quad \text{for} \quad n>0 \, .
\end{align}

\subsubsection{Discrete spectrum at the first dynode}

\begin{figure}
    \centering
    \includegraphics[width=0.7\columnwidth]{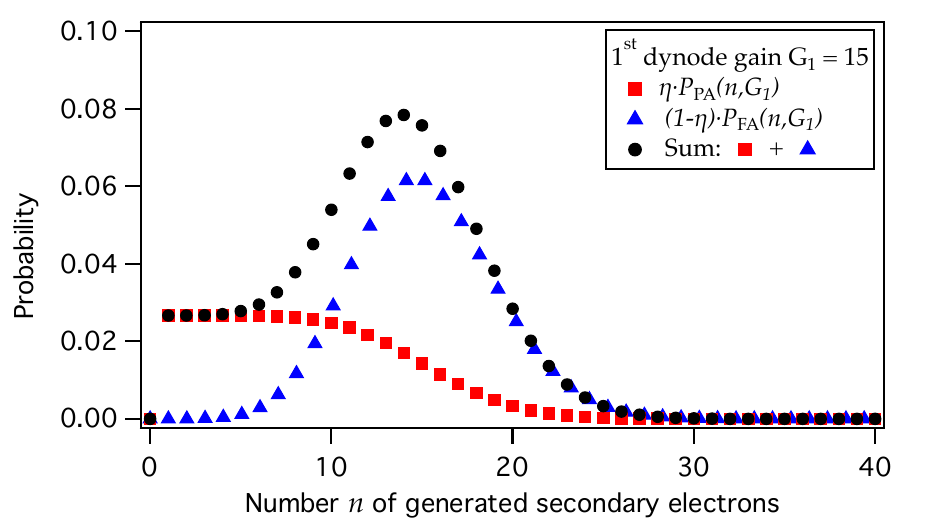}
    \caption{Black spots: Total discrete probability density for detectable secondary electrons emitted  at the first dynode. Blue triangles: contribution from fully amplified electrons. Red squares: contribution from partially amplified electrons. Here $\eta = 0.3$ and $G_1=15$.}
    \label{fig:discrete}
\end{figure}

The discrete probability density distribution for $n$ detectable secondary electrons emitted at the first dynode is given by the weighted sum of the contributions in Eqs.~\eqref{eq:Poisson} and \eqref{eq:cassetta_norm}:  
\begin{align}
\label{eq:P_1}
    P_{\rm 1{\rm Dy}}(n,G_1) = \left(1-\eta\right)P_{\rm FA}(n,G_1)+\eta P_{\rm PA} (n,G_1)\, .
\end{align}
It is completely described by the parameters $\eta$ and $G_1$. 
Figure~\ref{fig:discrete} shows the distributions of $P_{\rm FA}(n,G_1)$, in blue, and $P_{\rm PA}(n,G_1)$, in red, for $\eta = 0.3$ and $G_1 = 15$. In black is the weighted sum $P_{\rm 1{\rm Dy}}(n,G_1)$.

Clearly the mean gain $\langle G_1\rangle $ at the first dynode is smaller than $G_1$ due to the presence of the partially amplified primary photoelectrons.

\subsection{Effect of the successive dynodes}
\label{sec:succ_dynodes}
The amplification produced by the subsequent 
\mbox{$N-1$} dynodes smooths the discrete probability distribution described by Eq.~\eqref{eq:P_1}, rendering it continuous. 
In this section we derive the resulting analytical function, which features a very limited number of parameters. 
In the computation, we assume that each of these dynodes contributes with a mean gain $\langle G_i\rangle$, given by both fully and partially amplified electrons, and a variance $\sigma^2_i$. The overall gain $g_{N-1}$ and the resolution squared $R^2$ due to the $N-1$ dynodes after the first are then \cite{Wright}: 

\begin{align}
\begin{split}
&g_{N-1} = \prod_{i=2}^N \langle G_i\rangle,
\\
&R^2 = \frac{\sigma_2^2}{\langle G_2\rangle^2}+\frac{\sigma_3^2}{\langle G_2\rangle\cdot \langle G_3\rangle^2}+...+\frac{\sigma_N^2}{\langle G_2\rangle\cdot \langle G_3\rangle\cdot ...\cdot \langle G_N\rangle^2}.
\end{split}
\end{align}
The amplified signal is typically read out by front-end amplifiers and ADCs whose electronics contributes with an external noise that we assume to be Gaussian.
We define $f$ as the gain $g_{N-1}$ in ADC units of the \mbox{$N-1$} dynodes and $\sigma_{\rm ped}$ as the standard deviation of the readout noise, also in ADC units. 

Let us discuss the two terms on the right hand side of Eq. \eqref{eq:P_1} separately.

\subsubsection{Fully amplified continuous Poisson distribution}
Primary photoelectrons, fully amplified at the first dynode with gain $G_1$ and subsequently passing through the remaining $N-1$ dynodes with gain $g_{N-1}$, will yield an overall gain of $g = G_1\cdot g_{N-1}$, typically in the range $10^5-10^{7}$. 
The resulting continuous distribution, deriving from the convolution of the discrete function $P_{\rm FA}(n,G_1)$ with a Gaussian centred in $n$ and standard deviation proportional to $\sqrt{n}$ (Eq.~\eqref{eq:gauss}), is well described by a scaled continuous Poisson distribution~\cite{Ilienko,MARSAGLIA19861187} under the condition that $R^2G_1\gtrsim 1$ (see Appendix \ref{sec:appendix1}) as
\begin{align}
 {\mathscr P}_{\rm FA}(x) = \rho\frac{e^{-\rho\mu}\left(\rho\mu\right)^{\rho x}}{\Gamma(1+\rho x)},
\label{eq:cont_poisson1}
\end{align}
where the scaling factor $\rho$ is given by
\begin{align}
\rho = \frac{G_1f}{G_1f^2\left(1+R^2\right)+\sigma^2_{\rm ped}}\, .
\end{align}
The distribution only depends on four parameters: 
$G_1$, $f$, $R^2$ and $\sigma_{\rm ped}$.
The mean $\langle {\mathscr P}_{\rm FA}\rangle$ and the variance $\sigma_{{\mathscr P}_{\rm FA}}^2$ of the distribution are given by: 
\begin{align}
\begin{split}
\langle {\mathscr P}_{\rm FA}\rangle& = \mu = G_1 f ,\\
\sigma_{{\mathscr P}_{\rm FA}}^2 & = \frac{\mu}{\rho} = G_1f^2\left(1+R^2\right)+\sigma^2_{\rm ped}\,.
\end{split}
\label{eq:sigma_FA}
\end{align}

\subsubsection{Partially amplified continuous distribution}
The normalized discrete probability distribution  that characterizes partially amplified primary electrons at the first dynode (depicted by the red points in Figure~\ref{fig:discrete} and expressed in Eq. \eqref{eq:cassetta}) is  well described by the discrete  function
\begin{align}
\label{eq:g(x)}
    P_{\rm PA} (n,G_1) \propto 
    \left[1-{\rm erf}\left(\frac{n-G_1+\mu_0}{\sqrt{2G_1}}\right)\right]
\end{align}
for $G_1 \gtrsim 5$. A series of fits of Eq. \eqref{eq:cassetta} with \eqref{eq:g(x)} for various values of $G_1$ have shown that the falling edge is indeed directly related to $\sqrt{G_1}$ and that $\mu_0$ is a constant independent of $G_1$, with $\mu_0 = -0.62$ (see  Appendix \ref{sec:appendix2} for details).
 
 The amplification from the second dynode on, introduces a Gaussian smoothing centred in $n$ with standard deviation proportional to $\sqrt{n}$, being $n$ the number of secondary electrons  generated at the first dynode (see the details in Appendix \ref{sec:appendix2}). 
As a result of this convolution, the normalized distribution can be described by a rounded `box' shaped function
\begin{align}
\label{eq:cassetta_conv}
    {\mathscr P}_{\rm PA}(x) = \frac{\left[1+{\rm erf}\left(\frac{x-\mu_L}{\sqrt{2}\sigma_L}\right)\right]\left[1-{\rm erf}\left(\frac{x-\mu_R}{\sqrt{2}\sigma_R}\right)\right]}{4(\mu_R-\mu_L)}\, ,
\end{align} 
where both the rising and the falling edges of the distribution are well represented by ${\rm erf}(x)$ functions. 
The parameters $\sigma_L^2$ and $\sigma_R^2$, that describe the rate of rise and fall of the distribution, are the variances of the smoothing Gaussian for $n=1$ and $n=G_1$, respectively, and are derived in Appendix~\ref{sec:appendix2}. 
These can be expressed as: 
\begin{align}
\label{eq:sigmaL}
\sigma_L^2 &= {f^2R^2+\sigma_{\rm ped}^2},\\
\label{eq:sigmaR}
\sigma_R^2 &=  {f^2G_1\left(1+R^2\right)+\sigma_{\rm ped}^2}.
\end{align}
The parameters $\mu_L$ and $\mu_R$, derived in Appendix \ref{sec:appendix2}, are given by
\begin{align}
\label{eq:muL}
\mu_L &= f\left(0.5-0.45R^{2.2}\right),\\
\label{eq:muR}
\mu_R &= f\left(G_1+\mu_0-AR^{\gamma}\right),
\end{align}
with the parameters $\mu_0 = -0.62$ (same parameter as in Eq.~\eqref{eq:g(x)}), $A = 0.63$ and $\gamma = 1.7$, that are independent of $G_1$ and $R$. 

The mean and variance of ${\mathscr P}_{\rm PA}(x)$ can be calculated analytically and are given by: 
\begin{align}
\begin{split}
\langle {\mathscr P}_{\rm PA}\rangle  &= \frac{\mu_R+\mu_L}{2}\left(1+\frac{\sigma_R^2-\sigma_L^2}{\mu_R^2-\mu_L^2}\right),\\
\sigma^2_{{\mathscr P}_{\rm PA}} &= \frac{3\sigma^2_R\mu_R+\mu_R^3-3\sigma^2_L\mu_L-\mu_L^3}{3\left(\mu_R-\mu_L\right)}-\langle {\mathscr P}_{\rm PA}(x)\rangle ^2 \, . 
\end{split}
\label{eq:cassetta_mean_std}
\end{align}
Also in this case the distribution ${\mathscr P}_{\rm PA}$ only depends on the same four parameters $G_1$, $R$, $f$ and $\sigma_{\rm ped}$ as for ${\mathscr P}_{\rm FA}$.

\subsubsection{Additional contributions due to very low charge signals}
In addition to the contributions discussed in the previous sections, other multiplication effects occurring at the various dynodes can generate very low charge signals that emerge in the SPE response spectrum in the case the electronic noise is small enough, $\sigma_{\rm ped}\ll f$.
Among such effects one finds:
\begin{itemize}
\item photons that directly reach the first dynode without being absorbed on the photocathode.
These events appear as pre-pulses in a good time-resolution apparatus, being anticipated with respect to the normal signals by the electron transit time from the photocathode to the first dynode.
They generate a peaked distribution in the SPE response spectrum with a signal charge about a factor $G_1$ smaller than the fully amplified peak. They are also described by a continuous scaled Poisson distribution similar to that of Eq.~\eqref{eq:cont_poisson1} given by
\begin{align}
\label{eq:Pre-pulse}
{\mathscr P}_{\gamma\rm 1Dy}(x) = \rho'\frac{e^{-\rho'f'}\left(\rho'f'\right)^{\rho' x}}{\Gamma(1+\rho' x)},
\end{align}
with parameters $f'$ and $\rho'$ that account for the reduced total gain and modified resolution $R'$:
\begin{align}
\begin{split}
f' &= f \cdot G_2 / \langle G_2 \rangle = f\cdot \zeta  , \\\rho'&=\frac{f'}{f'^2R'^2+\sigma_{\rm ped}^2}\, .
\end{split}
\end{align}
Here, $\zeta$ is the ratio between the full gain at the second dynode, $G_2$, and its average value (including back-scattering and other processes) $\langle G_2 \rangle$ and (see Appendix \ref{sec:appendix_prepulse})
\begin{align}
    R'^2 \approx \frac{R^2}{\zeta}\left[1- \frac{\zeta-1}{R^2+1}\right].
\end{align}
Similarly to Eq.~\eqref{eq:sigma_FA}, the mean and variance of ${\mathscr P}_{\gamma \rm 1Dy}(x)$ are given by:
\begin{align}
\begin{split}
\langle{\mathscr P}_{\gamma \rm 1Dy}\rangle &= f' ,\\
\sigma^2_{{\mathscr P}_{\gamma \rm 1Dy}} &= 
    \frac{f'}{\rho'}= f'^2R'^2+\sigma^2_{\rm ped} \, .
\end{split}
\label{eq:R'}
\end{align}
Notice that only the new parameter $\zeta$ has been introduced to describe the pre-pulse contribution in Eq.~\eqref{eq:Pre-pulse}.

\item photoelectrons generated at the photocathode which miss the first and possibly other dynodes (bad trajectory). 
This  contribution can be described by an exponentially modified Gaussian, centred at the origin with standard deviation $\sigma_{\rm ped}$ and with decay constant $\alpha$:
\begin{align}
\label{eq:low-charge}
    {\mathscr P}_{\rm Exp}(x) = \frac{1}{2\alpha}e^{\frac{1}{2\alpha}\left(\frac{\sigma_{\rm ped}^2}{\alpha}-2x\right)}{ \rm erfc}\left(\frac{\frac{\sigma_{\rm ped}^2}{\alpha} - x}{\sqrt{2}\sigma_{\rm ped}}\right).
\end{align}
The steepness of the cut-off towards zero charge is fixed by $\sigma_{\rm ped}$, whereas the parameter $\alpha$ must be $\alpha\sim f$ given that on average the amplification of such signals can be at most $f$. In fact, signals with amplitudes greater than $f$ must have originated at the first dynode.

The mean and variance of ${\mathscr P}_{\rm Exp}(x)$ are:
\begin{align}
\begin{split}
    \langle{\mathscr P}_{\rm Exp}\rangle &= \alpha,\\
    \sigma^2_{{\mathscr P}_{\rm Exp}} &= \alpha^2+\sigma_{\rm ped}^2\, .
\end{split}
\end{align}
\end{itemize}
The inclusion of such additional contributions in the description of the SPE response function is of particular importance when a precise acceptance of single photoelectrons is of interest.

\subsubsection{Analytical spectral response function}\label{sec:model}
The complete SPE response model that accounts for the processes discussed so far is given by 
\begin{align}
\begin{split}
    {\mathscr P}_{\rm SPE}(x) &= (1-\eta-A_{\gamma \rm 1Dy}-A_{\rm Exp}){\mathscr P}_{\rm FA}(x)+\\&+\eta {\mathscr P}_{\rm PA}(x)+A_{\gamma \rm 1Dy}{\mathscr P}_{\gamma \rm 1Dy}(x)+A_{\rm Exp}{\mathscr P}_{\rm Exp}(x)\, ,
\end{split}
\end{align}
where $\eta$, $A_{\gamma  \rm 1Dy}$, $A_{\rm Exp}$ and $(1-\eta-A_{\gamma \rm 1Dy}-A_{\rm Exp})$ are the fractions of the four probability density functions ${\mathscr P}_{\rm FA}(x)$, ${\mathscr P}_{\rm PA}(x)$,  ${\mathscr P}_{\gamma \rm 1Dy}(x)$ and ${\mathscr P}_{\rm Exp}(x)$, 
respectively.
Notice that ${\mathscr P}_{\rm SPE}(x)$ depends on a limited number of free parameters. Besides the three fractions, the other free parameters are $G_1$, $\alpha$ and $R$ describing the photomultiplier response, $\sigma_{\rm ped}$ describing the readout electronics, $f$ depending on both the photomultiplier response and on the readout electronics, and $\zeta$ representing the ratio between the full gain at the second dynode and its average value. 

A typical light spectrum is obtained by calibrating a photomultiplier with a triggered, very-low intensity, sub-nanosecond light pulse with an occupancy $\cal Q$ below about $20\%$. These calibration spectra will also contain two and more PE contributions. 
Based on the central limit theorem, to describe these contributions we approximate the convolution of two or more single electron response spectra with a Gaussian function whose parameters are related to the SPE response. Although the central limit theorem works better for a large number of photoelectrons, this approximation is sufficiently good given that the fraction of two photoelectron events is suppressed by a factor ${\cal Q}/2$ (assuming Poisson statistics). 

The mean and standard deviation of the SPE response can be determined analytically from the function ${\mathscr P}_{\rm SPE}(x)$:
\begin{align}
\begin{split}
    \langle {\mathscr P}_{\rm SPE}\rangle &= \left(1-\eta-A_{\gamma \rm 1Dy}-A_{\rm Exp}\right)\mu+\eta\langle {\mathscr P}_{\rm PA}\rangle
    +A_{\gamma \rm 1Dy}f +A_{\rm Exp}\alpha,
    \\
       \sigma^2_{\rm {\mathscr P}_{\rm SPE}} &= (1-\eta-A_{\gamma \rm 1Dy}-A_{\rm Exp})\left[\sigma^2_{\rm {\mathscr P}_{\rm FA}}+\mu^2\right]+ 
       \eta \left[\sigma^2_{\rm {\mathscr P}_{\rm PA}}+\langle {\mathscr P}_{\rm PA}\rangle^2\right] + \\ 
       &+ A_{\gamma \rm 1Dy}\left[\sigma^2_{{\mathscr P}_{\gamma \rm 1Dy}}+f^2\right]+
        A_{\rm Exp}\left[\sigma^2_{{\mathscr P}_{\rm Exp}}+\alpha^2\right]
        -\langle {\mathscr P}_{\rm SPE}\rangle^2\, .
\end{split}
\end{align}
The photoelectron convolution response for multiple photoelectrons, $M$, will therefore have mean $M\langle {\mathscr P}_{\rm SPE}\rangle $. Given that the contribution of the ADC noise comes after the charge collection at the anode, to determine the variance of the convolution we first subtract $\sigma^2_{\rm ped}$ from $\sigma^2_{{\mathscr P}_{\rm SPE}}$ before multiplying by $M$. Therefore, $\sigma^2_{M}= M\left(\sigma^2_{{\mathscr P}_{\rm SPE}}-\sigma^2_{\rm ped}\right)+\sigma^2_{\rm ped}$.
Here we have made the approximation in Eq. \eqref{eq:cassetta_mean_std} that $\sigma^2_{\rm {\mathscr P}_{\rm PA}}+\langle {\mathscr P}_{\rm PA}\rangle^2$ contains an additive  $\sigma^2_{\rm ped}$ term.

\section{Experimental setup}\label{sec:setup}
\begin{figure*}[tb!]
    \centering
    \includegraphics[width=0.6\textwidth]{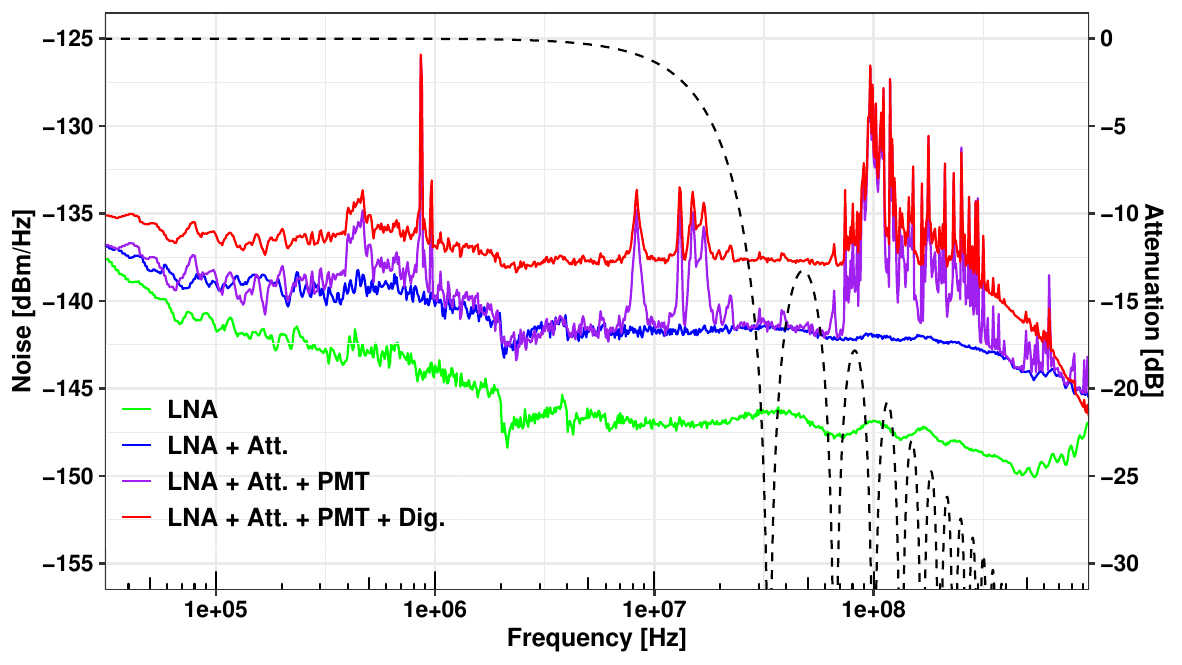}
    \caption{Noise spectra of the different components of the read-out spectra. All spectra are referred to the input of the digitizer in power density per Hertz. The feature at 2--3~MHz is instrumental from the Rohde \& Schwarz FPL1003 spectrum analyser. Also note the pick-up noise structure at frequencies peaking at 100~MHz. The attenuation introduced by the waveform integration for 30~ns is shown in black (referred to the right scale). This plot refers to the photomultiplier Hamamatsu R5912-100.}
    \label{fig:noise}
\end{figure*}

The analytical model described in the previous section has been verified using two different models of photomultipliers: an 8-inch Hamamatsu R5912-100 (used, e.g. in the XENONnT~\cite{nVeto} and LZ~\cite{LZ} neutron vetoes) and a 3-inch Hamamatsu 6233. The setup was prepared at the LNGS external facilities. The photomultipliers under test were installed in a dark box equipped with optical and electrical feed-through. The box and the connections were further shielded by a black cloth to prevent residual ambient photons from interacting with the device under study.

A low noise HV power supply (ISEG SHQ) was used to provide the bias for the photomultiplier. The signal extracted from the anode was amplified by the low-noise amplifier (RFBAY LNA-150) and then forwarded to a high speed digitizer. The low-noise amplifier (LNA) provides an amplification of a factor 42~V/V in the frequency range 5~kHz to 300~MHz. The signal is acquired by a fast digitizer (CAEN V1751) with a sample rate of 2~GS/s and a depth of 10~bits on a input range of 1~V. 
Since the input impedance of the LNA is not perfectly matched at 50~$\Omega$ and given that the photomultiplier base is not back-side terminated, a passive attenuator of 6~dB was installed in front of the amplifier to avoid signal reflections.

The readout chain was selected to acquire waveforms at high speed with low electronic noise: the LNA exhibits a very low noise figure (${\rm NF}=2$) equivalent to 300~$\mu$V (ac-RMS) at the output of the amplifier on the full bandwidth. However the Johnson–Nyquist thermal noise of the 6~dB attenuator has to be accounted as well (with ${\rm NF}=3$). As such the noise after the amplifier reaches 500~$\mu$V, lower than the input noise of the digitizer (600~$\mu$V).
Overall the white noise of the system is about 800~$\mu$V (at one sigma) corresponding to a charge noise of $4\times 10^4\, e^-$ (integration time 30~ns) at the anode of the photomultiplier, where typical signals are between $5\times 10^5\, e^-$ and $1\times 10^7\, e^-$. 
Figure~\ref{fig:noise} shows the noise spectrum of the read-out chain, with the different contributions highlighted. Beyond the intrinsic stationary processes of the readout chain, photomultipliers may introduce pick-up noise. Since most of these external contributions are at high frequency, their repercussions to the integrated waveform is strongly attenuated.

Pico-second lasers (Edinst EPL405 and EPL510) were used to deliver short light pulses (below 100~ps) 
at $\lambda=\,$405~nm and 507~nm.  
The laser sync signal was used to trigger the digitizer for a symmetric window of 500~ns. 
A custom attenuation box allowed to reduce the power of the laser optical pulse in order to have a low photon occupancy per each trigger on the photomultiplier. 
The laser pulses were delivered at fixed positions on the photocathode via optical fibres. 


The events were stored on disk by a PC-based acquisition system for the offline analysis.
Samples of 600k events were acquired at different HV biases, illuminating positions and light wavelengths. Each measurement is complemented by a {\it dark} background sample acquired with the same setup, but putting a shutter on the light source. This sample is used in the offline analysis to subtract the pedestal and dark-count contributions from the spectrum  (see Section~\ref{sec:analysis}).
The measurements were performed at room temperature after waiting several hours after the HV was turned on and the dark box was closed. This was necessary to reach a reasonably low level (${\cal O}({\rm kHz}))$ of dark-rate counts.

\section{Analysis}\label{sec:analysis}

\begin{figure*}[h!]
    \centering
    \includegraphics[width=0.46\textwidth]{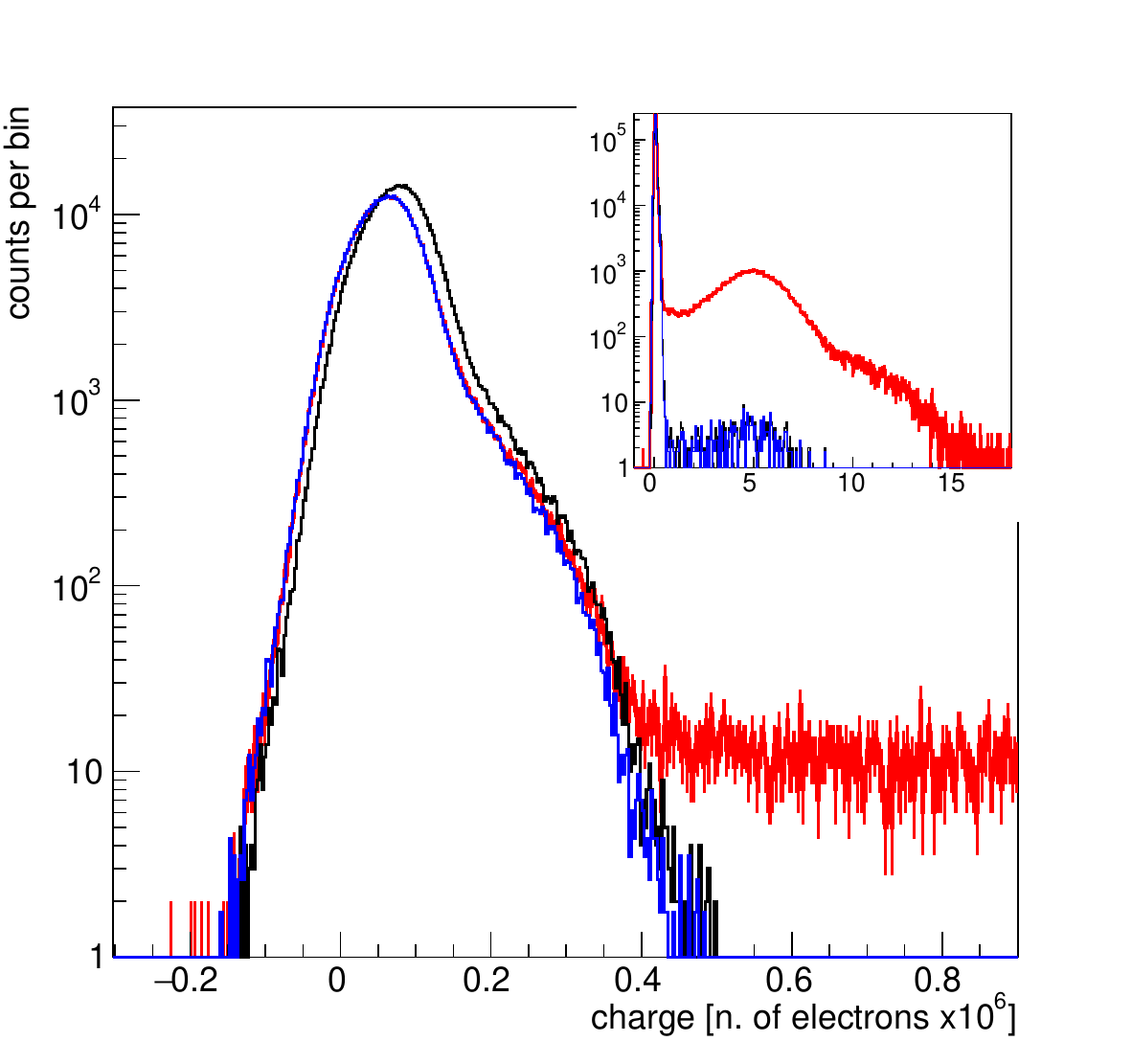}
    \includegraphics[width=0.46\textwidth]{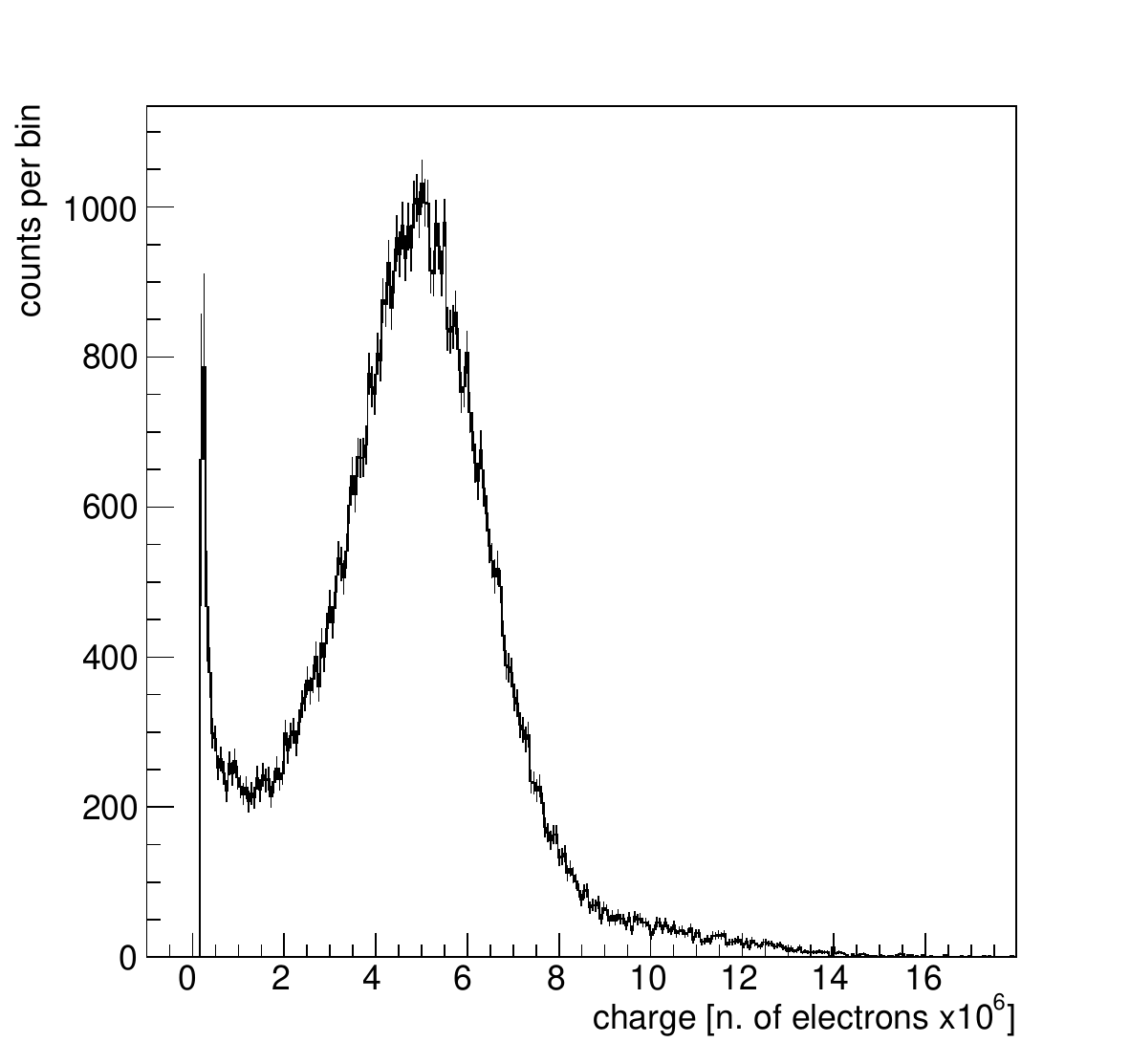}\\
    \caption{Left: Typical spectral response of Hamamatsu R5912-100 obtained with a light source at 405~nm (red histogram) and with light-off (black histogram). The \emph{light-only} spectrum is obtained by subtracting the \emph{dark}  histogram from the \emph{light-on} one after applying to the first one a global shift and a scaling factor to equalize the pedestal peak contributions (blue histogram). Right: subtracted spectral response.}
    \label{fig:spectrum-5912-100}
\end{figure*}
\begin{figure*}[h!]
    \centering    \includegraphics[width=0.46\textwidth]{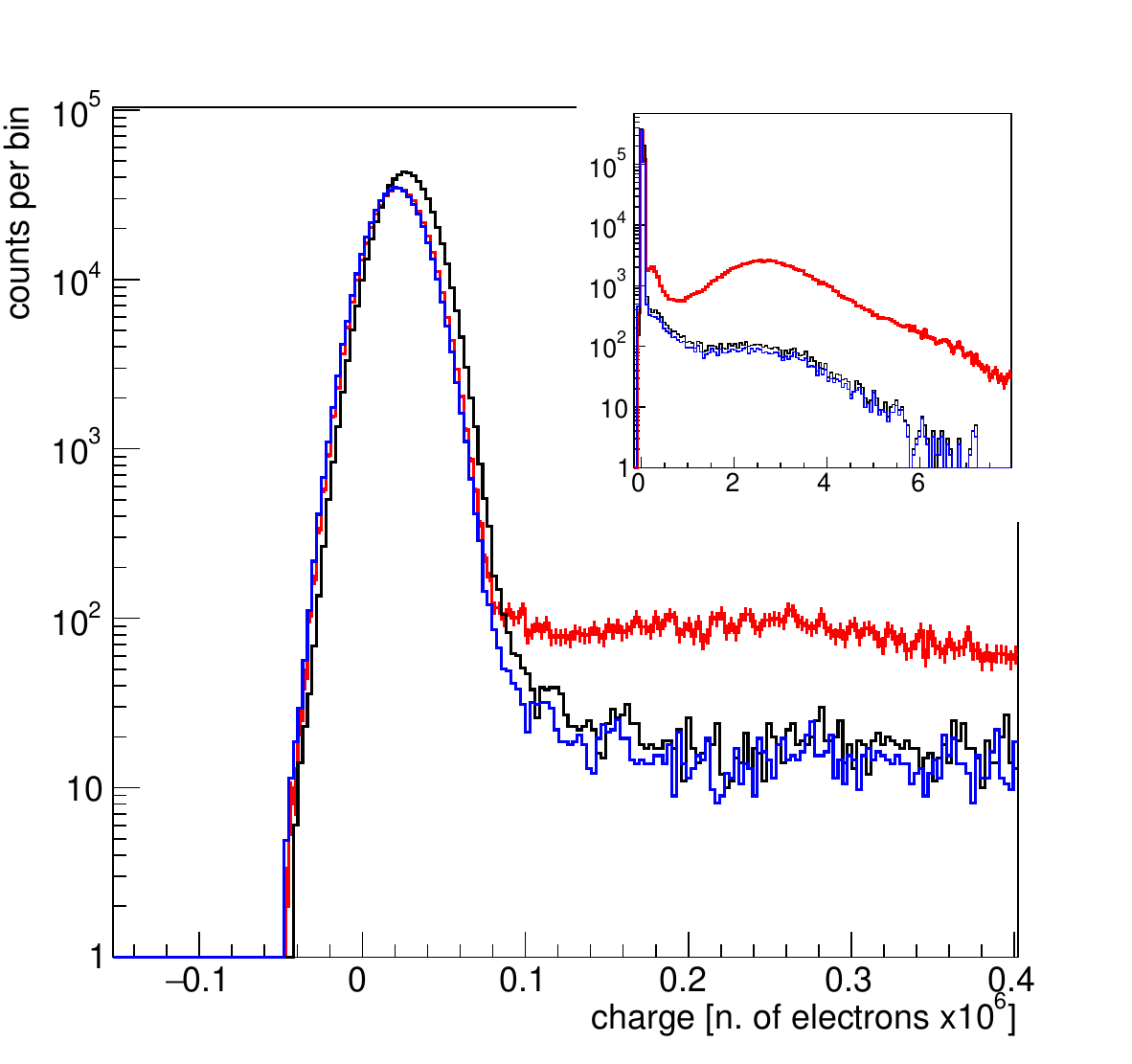}\includegraphics[width=0.46\textwidth]{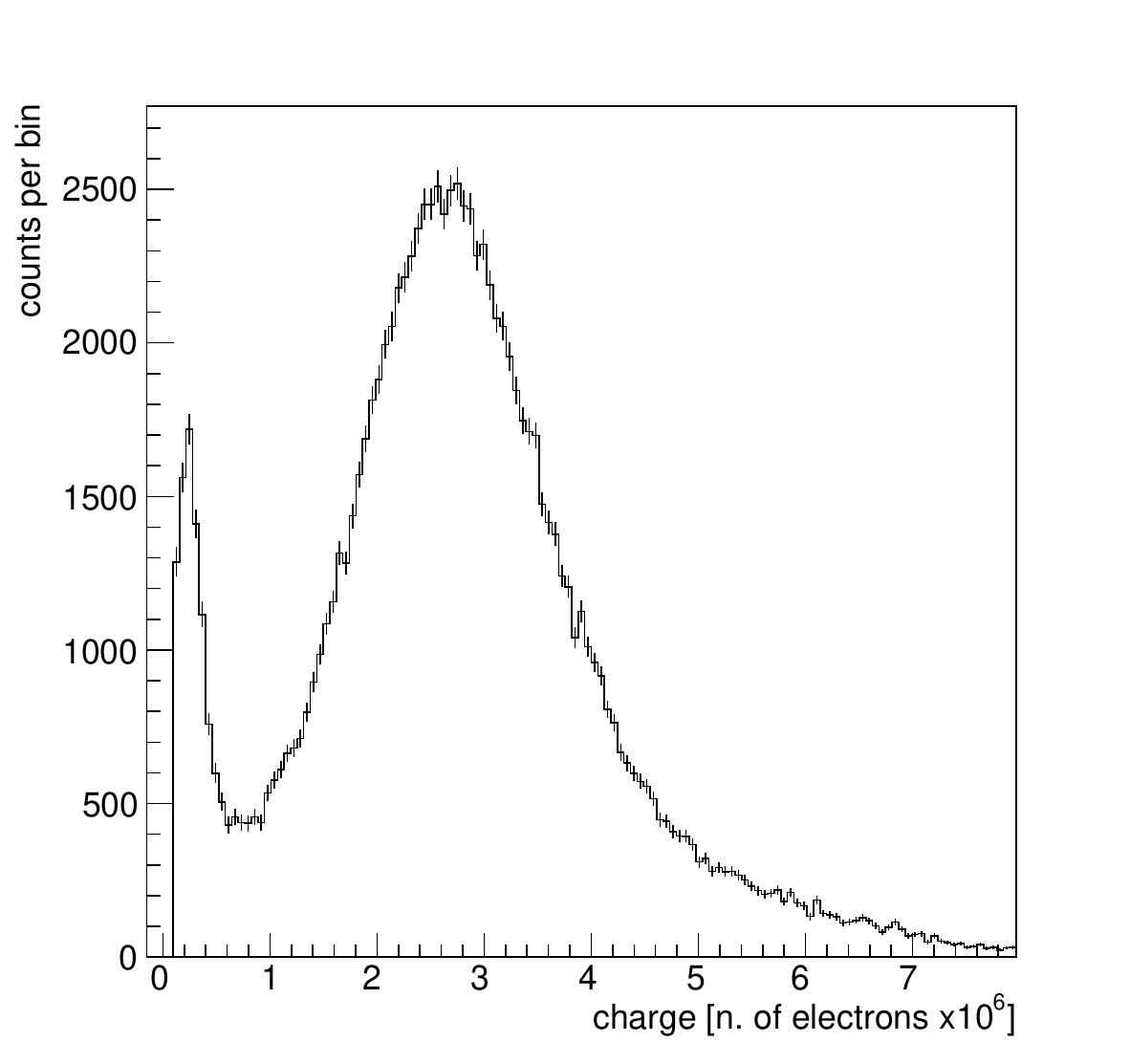}    \caption{Typical spectral response of Hamamatsu 6233. See Fig.~\ref{fig:spectrum-5912-100} caption. }
    \label{fig:spectrum-6233}
\end{figure*}

Each acquired waveform consists in 1000 samples (with 0.5~ns/sample). The signal due to the laser pulse starts at the sample n. 640. The first 50 samples of the waveform are used to determine the baseline. 

Each acquired waveform is subtracted from the mean baseline value and reversed to be a positive signal. 
The charge spectrum is obtained by searching for the sample $t_{\rm max}$ corresponding to the maximum ADC value in a window of \mbox{[-50,+75[~ns} after a time interval of 240~ns from the laser trigger signal for the Hamamatsu R5912-100 and 90 ns for the Hamamatsu 6233.
The waveform is then integrated in a time interval of \mbox{[-6,+24[} samples \mbox{([-3,+12[~ns)} from $t_{\rm max}$ for both the photomultipliers.

\begin{figure*}[!t]
    \centering
    \includegraphics[width=0.46\textwidth]{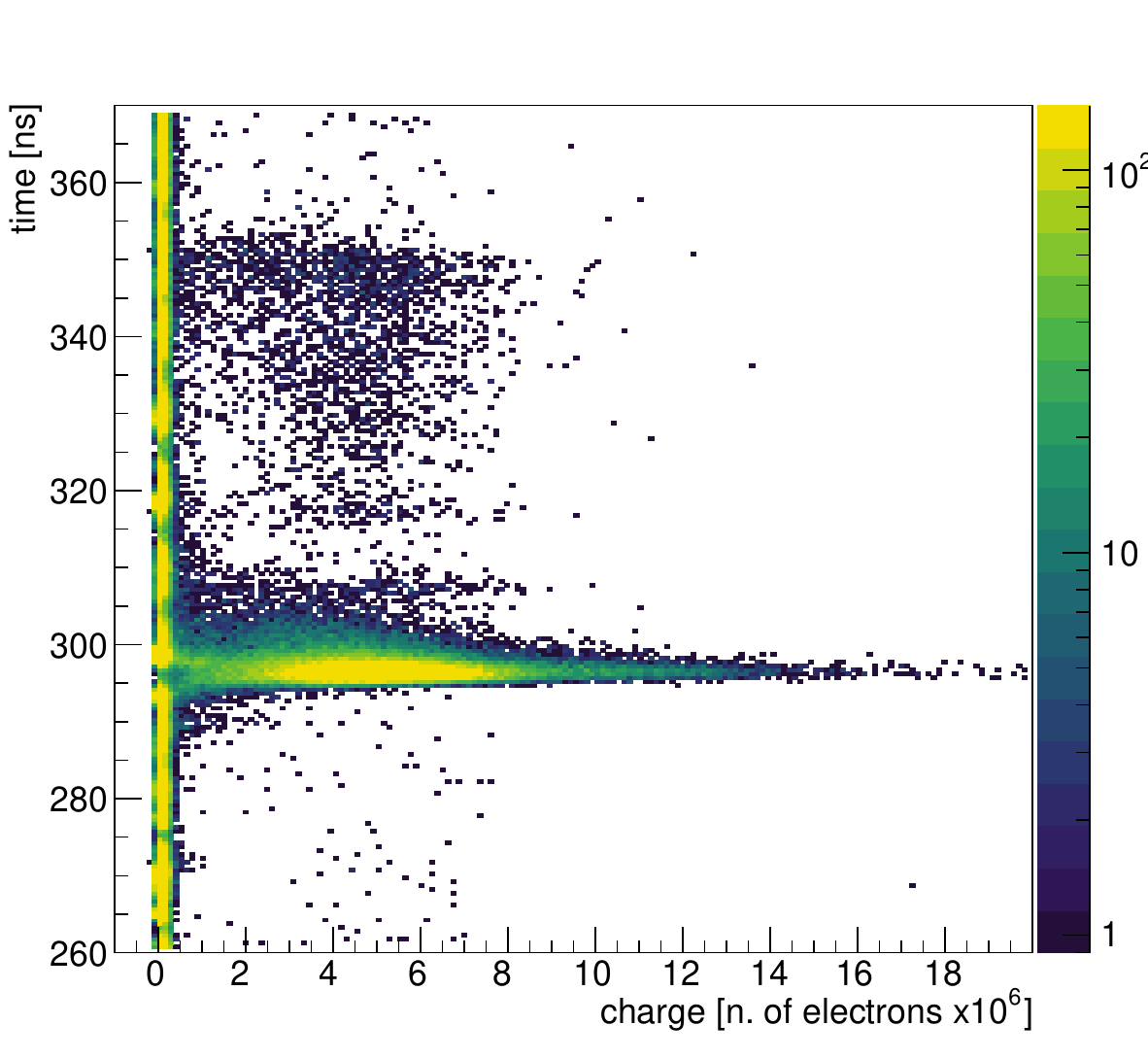}
    \includegraphics[width=0.46\textwidth]{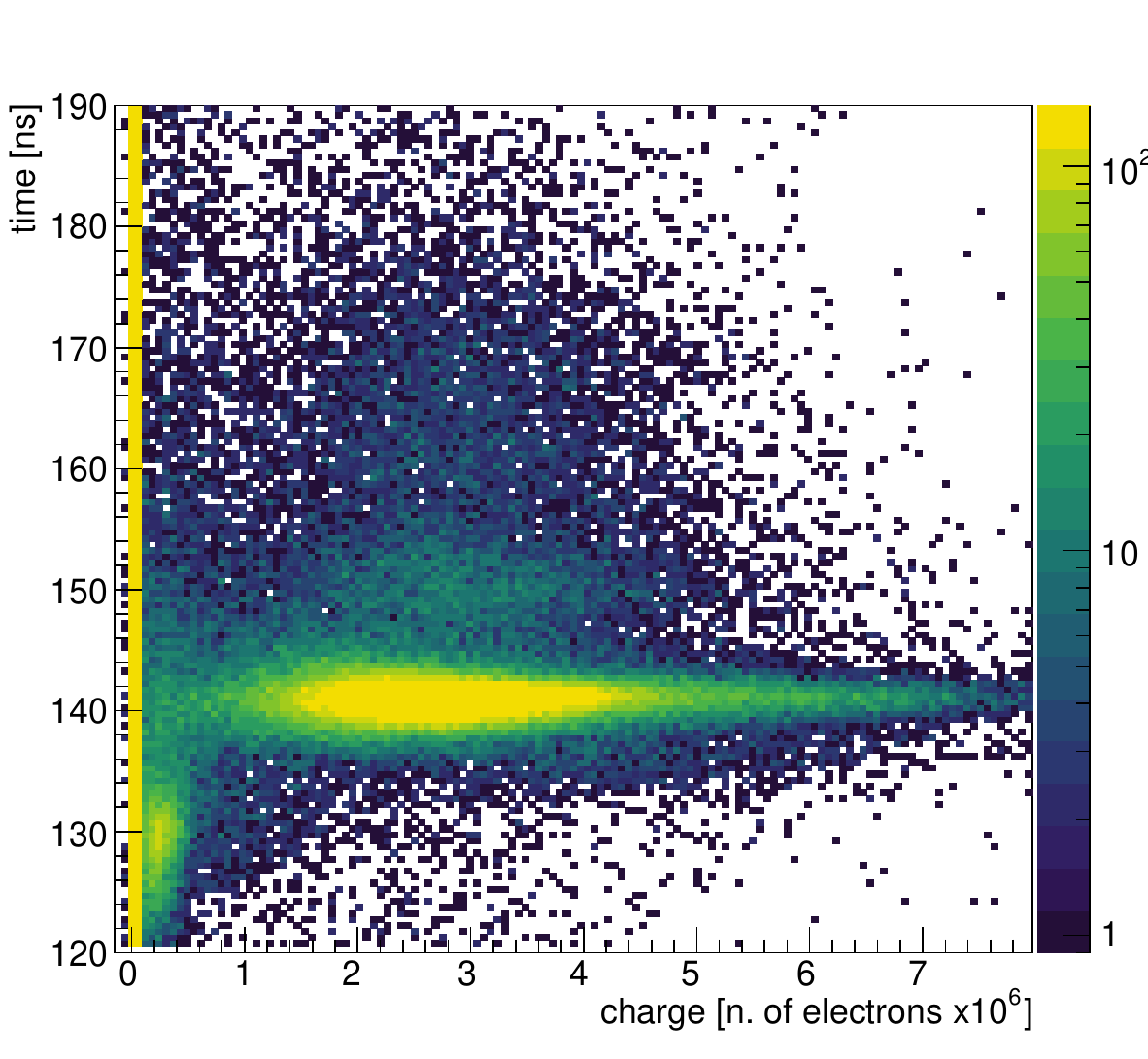}\\    
    \caption{Signal time vs spectral response for photomultiplier models (left) Hamamatsu R5912-100 and (right) Hamamatsu 6233. The vertical yellow band around zero charge is due to the pedestal. }
    \label{fig:ScatterPlot}
\end{figure*}

Figures~\ref{fig:spectrum-5912-100} and \ref{fig:spectrum-6233} show a typical spectrum with light on (red histogram in the insert) where the charge is expressed in units of number of electrons, calculated using the ADC counts, the ADC units and the amplification factor.
From left to right, are clearly visible a dominating very narrow pedestal peak around zero, a valley followed by a second wider peak, due to fully amplified single photoelectron signals, and finally by a decreasing structure due to two and more photoelectron events. 
The corresponding spectrum with light off (black histogram) from the \emph{dark} data sample shows the pedestal peak with the same features, though slightly shifted to larger values, and a small contribution of signals due to dark rate.  
The light-only spectrum is derived by subtracting the dark histogram from the light-on histogram after correcting the former both for a global offset and a counts-per-bin normalization factor in order to equalize the pedestal contributions (blue histogram). The offset is determined by scanning discrete bin shifts and selecting the value that minimizes a $\chi^2$ between the two spectra within a window of about  $ \pm 1\sigma_{\rm ped}$ around the pedestal peaks, assuming Poisson fluctuations. Once the optimal shift is found, the normalization factor is computed as the ratio of the integrals of the two spectra in a fixed window around the pedestal peaks.
As a consequence, the subtracted dark rate contribution turns out to be slightly underestimated. This effect has a negligible impact on the subtracted spectrum given the low dark count rate.

\begin{figure*}[t!]
    \centering
\includegraphics[width=0.46\textwidth]{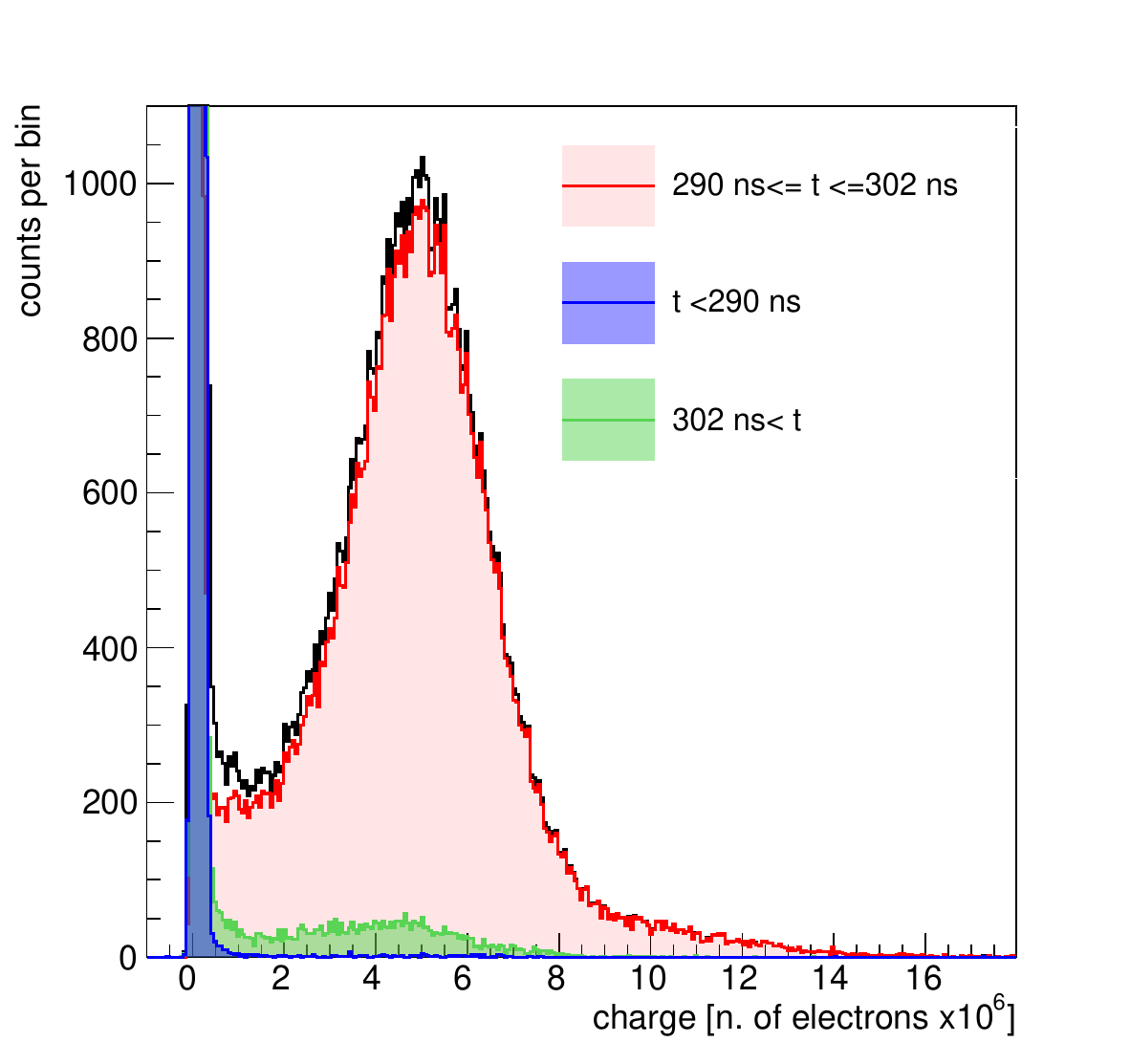}
    \includegraphics[width=0.46\textwidth]{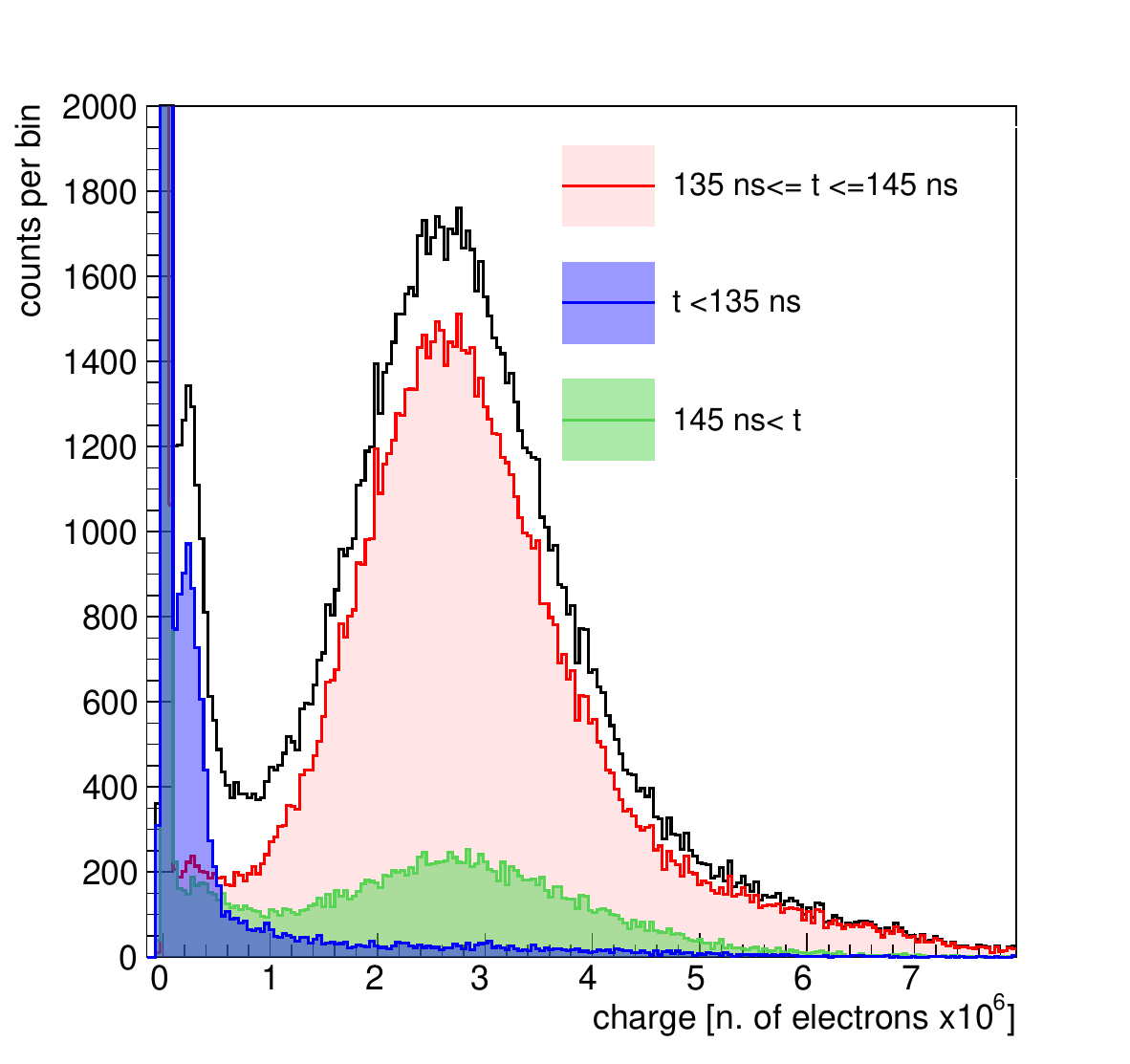}
    \caption{(Left) Spectral response in different time regions for  Hamamatsu R5912-100. The red, blue and green histograms refer to the on-time, delayed and anticipated events defined by the time windows [292; 300]~ns, $>$304~ns and $<$280~ns, respectively. The black histogram is the sum of the three histograms. (Right) Same plot for Hamamatsu 6233. In this case the respective time intervals are [136; 144]~ns, $>$147~ns and $<$132~ns. }
    \label{fig:SpettriVarieRegioni}
\end{figure*}

 Particularly interesting for the validation of the proposed model in Section~\ref{sec:section2} is the region of the spectrum below the single photoelectron peak which extends to very low values of the collected charge ($\sim 1\times10^{6} $ electrons) shown in Figures~\ref{fig:spectrum-5912-100} and \ref{fig:spectrum-6233}, right.\footnote{The inspection of this region benefits from using the low-noise amplifier that enhances the S/N ratio.} 
In this region, the light-only spectrum of both the photomultipliers show a flat distribution and a structure at very low charge that has an exponential shape for the Hamamatsu R5912-100 and a peaked shape for the Hamamatsu 6233 photomultipliers. 
By inspecting the correlations of the charge with the signals' time of arrival, shown in Figures~\ref{fig:ScatterPlot} and~\ref{fig:SpettriVarieRegioni}, it is possible to address these last two structures to different origins. 
In fact, while for the R5912-100 photomultiplier the events corresponding to the exponential shape have no clear time structure, for the  6233 photomultiplier the low charge events feature a peaked structure in time about 10~ns earlier than that of the typical photoelectron signals, associated to the most dense horizontal band at $\sim$141~ns. 
Such events are interpreted as pre-pulses, and are described by Eq.~\ref{eq:Pre-pulse}.
 
In addition to the clear excess of events due to single photoelectron signals accumulating around the signal time of arrival of 293~ns (140~ns) for R5912-100 (6233) photomultiplier, referred to as on-time, both the photomultipliers also show a small fraction of events delayed by 20--45~ns (10-15~ns) with a charge distribution similar to that of the on-time signals. Such events are due to back-scattered electrons on the first dynode which, if not lost, bounce back down the dynode chain completing the amplification with a slight delay.

Finally the pedestal region in Figure \ref{fig:ScatterPlot}, left, related to the photomultiplier R5912-100  shows a time structure. 
This is due to pick-up noise structure at frequencies peaking at 100~MHz visible in Figure~\ref{fig:noise}. A similar time structure is not present for the photomultiplier 6233 because it was mounted in a metal shielded box.

\begin{figure}[h!]
    \centering
\includegraphics[width=0.45\textwidth]{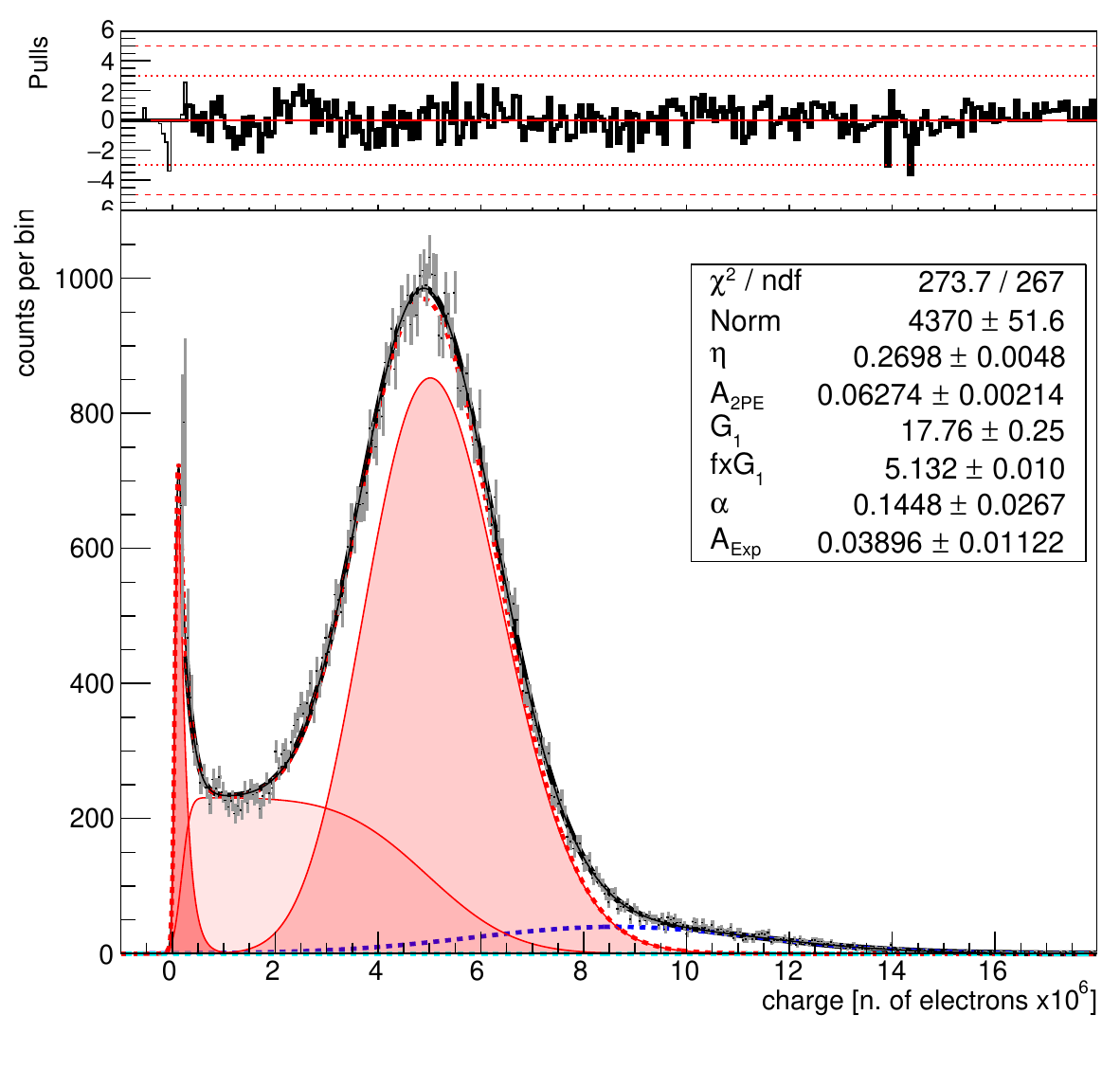}\\
\caption{Fit to the spectral response of light-only contribution for the R5912-100 photomultiplier corresponding to the measurement with HV=1200~V, illuminated with light wavelength of 405~nm  
with a fibre in central position. 
The fit result is shown by the thick black line.
The fit function consists of the contribution from  1PE (dashed red)  and 2PE (dashed blue). The 1PE contribution consists in the sum of three components: a low-charge (red filled area), described by Eq. ~\eqref{eq:low-charge}; a partially amplified (red light-shaded area), modelled by the function in Eq.~\eqref{eq:cassetta_conv}; and fully-amplified component (red medium-shaded filled area) corresponding to Eq.~\eqref{eq:cont_poisson1}. The 2PE contribution is described by a Gaussian function with parameters derived from the 1PE parameters. See Section~\ref{sec:analysis} for details. The top panel shows the normalized residuals (\emph{pulls}).}
    \label{fig:my_label}
\end{figure}

Each subtracted spectrum (Figures~\ref{fig:spectrum-5912-100} and \ref{fig:spectrum-6233}, right) is fit with the model discussed in Section~\ref{sec:model}. 
The free fit parameters are the  gain of the first dynode, $G_1$, the mean of the fully amplified photoelectrons, \mbox{$\mu \equiv G_1 \times f$},  the fractions within the 1PE contribution of the back-scattered photoelectrons, $\eta$, of the exponential contribution, $A_{\rm Exp}$, and of the pre-pulse, $A_{\gamma \rm 1Dy}$, when needed, as well as the exponential decay constant, $\alpha$ and the parameter $\zeta$. 
Additional parameters are the histogram normalization, $\rm Norm$, and the fraction of two and three photoelectrons, $A_{\rm 2PE}$ and $A_{\rm 3PE}$, if needed. 
 Finally, the parameters such as the mean and sigma of the pedestal, $\mu_{\rm ped}$ and $\sigma_{\rm ped}$, are fixed to the values determined from fits to the \emph{dark} data sample, shifted to match the signal pedestal distribution.
The parameter $R$ that contributes to the width of the partially and fully amplified distributions in the photomultiplier R5912-100 is weakly constrained by the data and highly correlated with $G_1$. 
For this reason it is fixed to a value determined from simulations and the resistive dynode chain ($R = 0.435$). On the contrary, for the photomultiplier 6233 the presence of the well separated pre-pulse peak allows the determination of $R$ (and also the parameter $\zeta$) which is thus a free parameter.
 
 The best fit parameters are obtained by minimizing the chi-square between the fit function and the histogram bin entries, and accounting for the bin uncertainties. 

\section{Results}
\begin{figure*}[!h]
    \centering
    \includegraphics[width=0.45\textwidth]{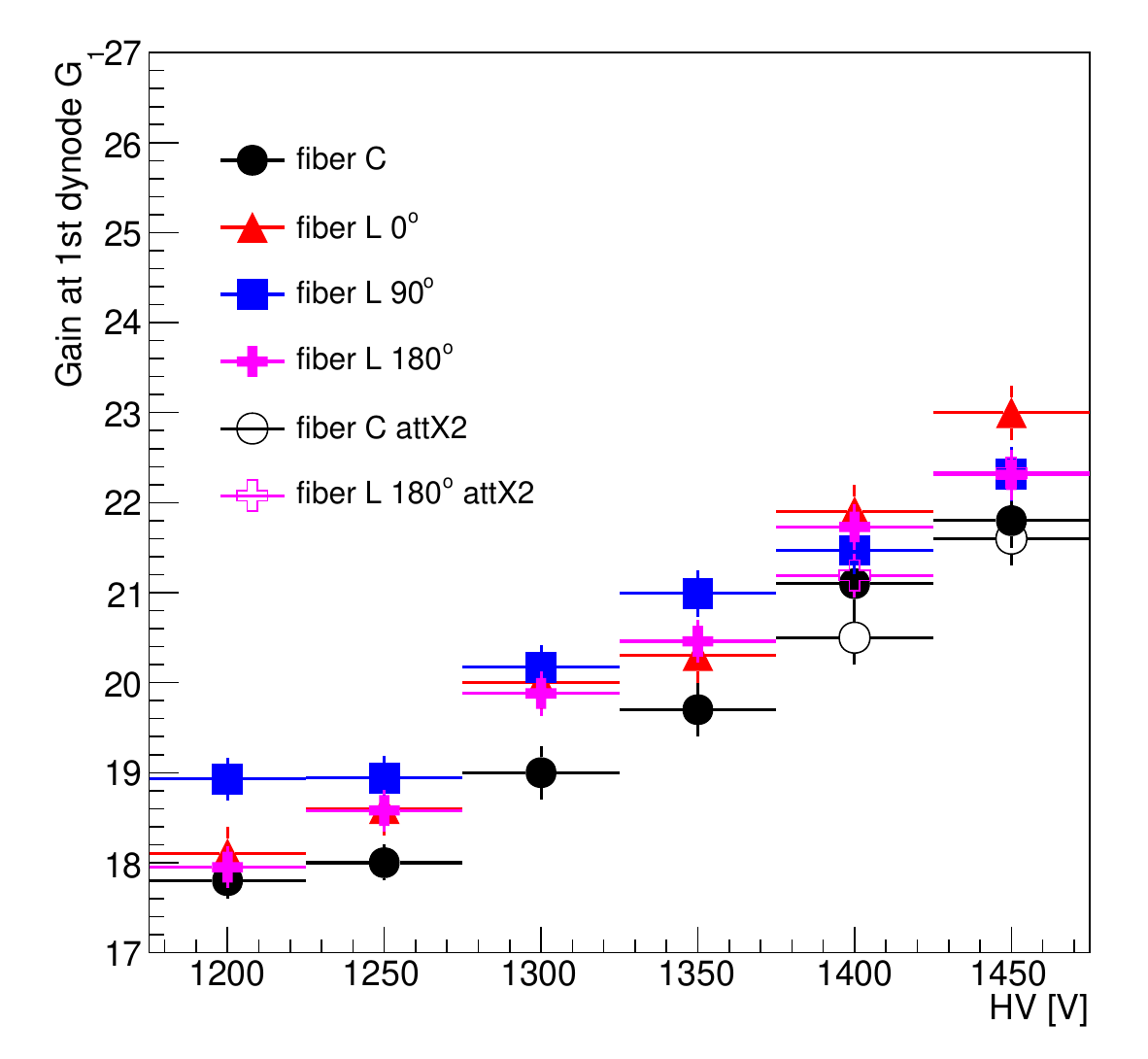}
    \includegraphics[width=0.45\textwidth]{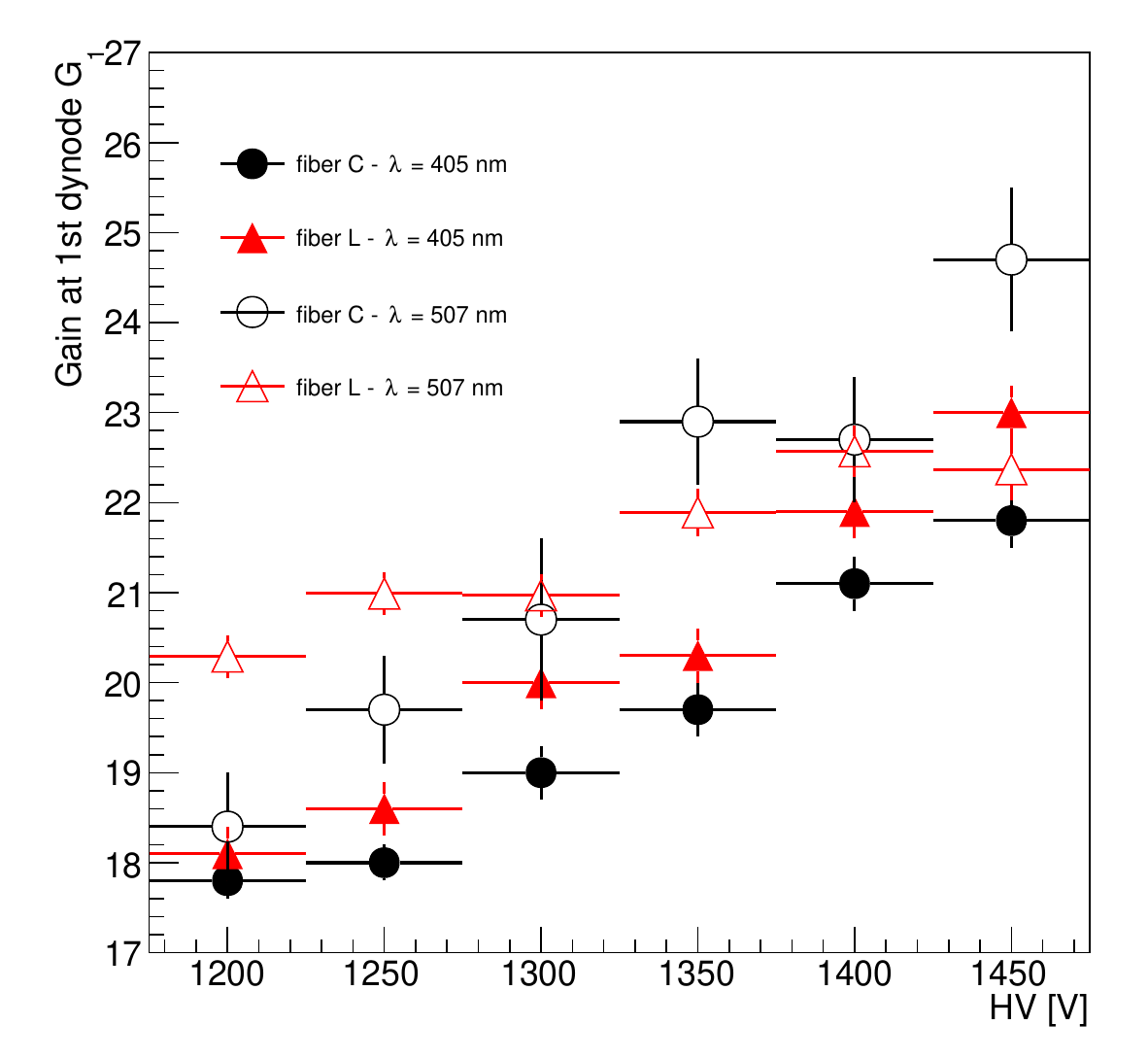}\\
    \includegraphics[width=0.45\textwidth]{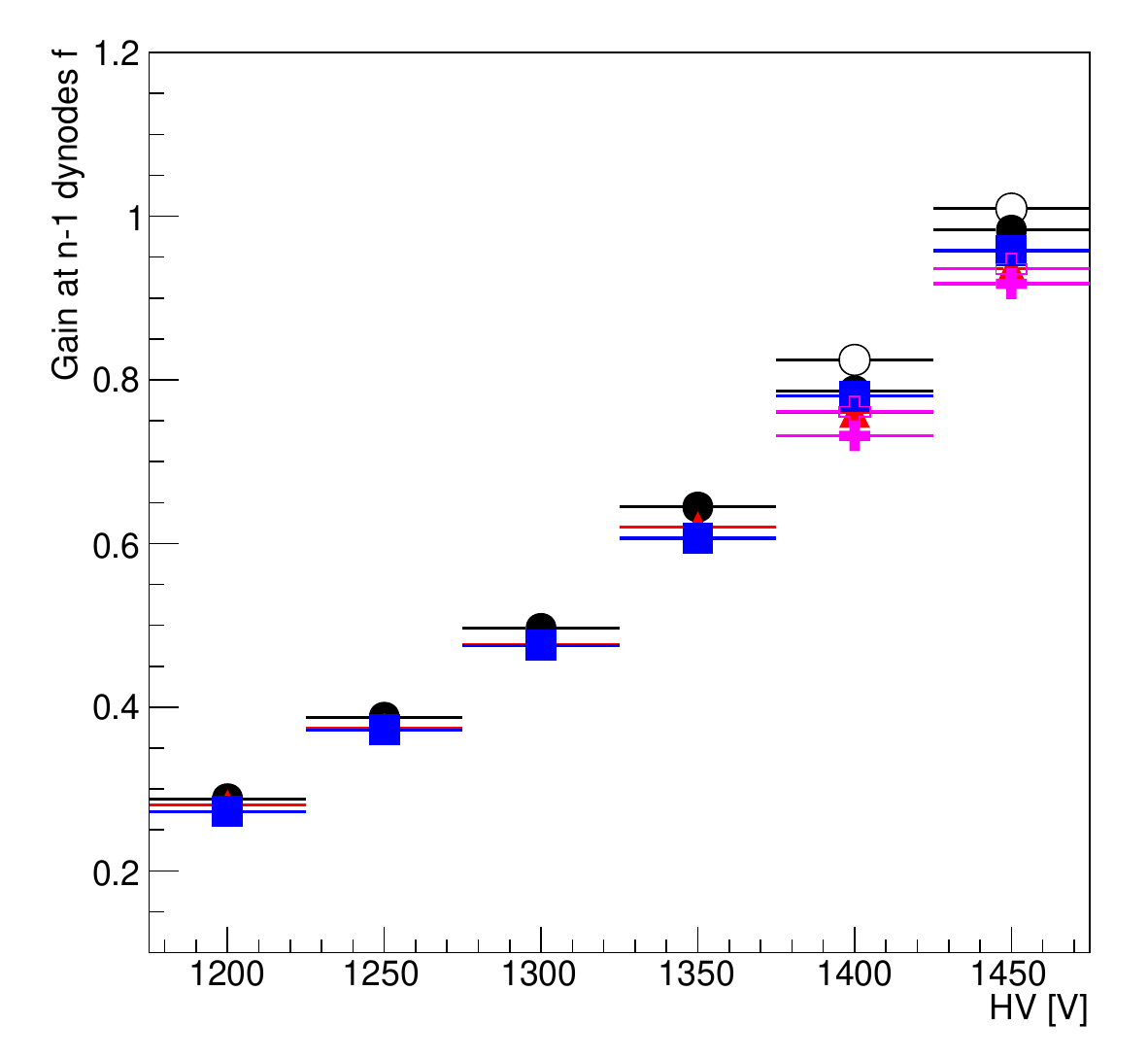}
    \includegraphics[width=0.45\textwidth]{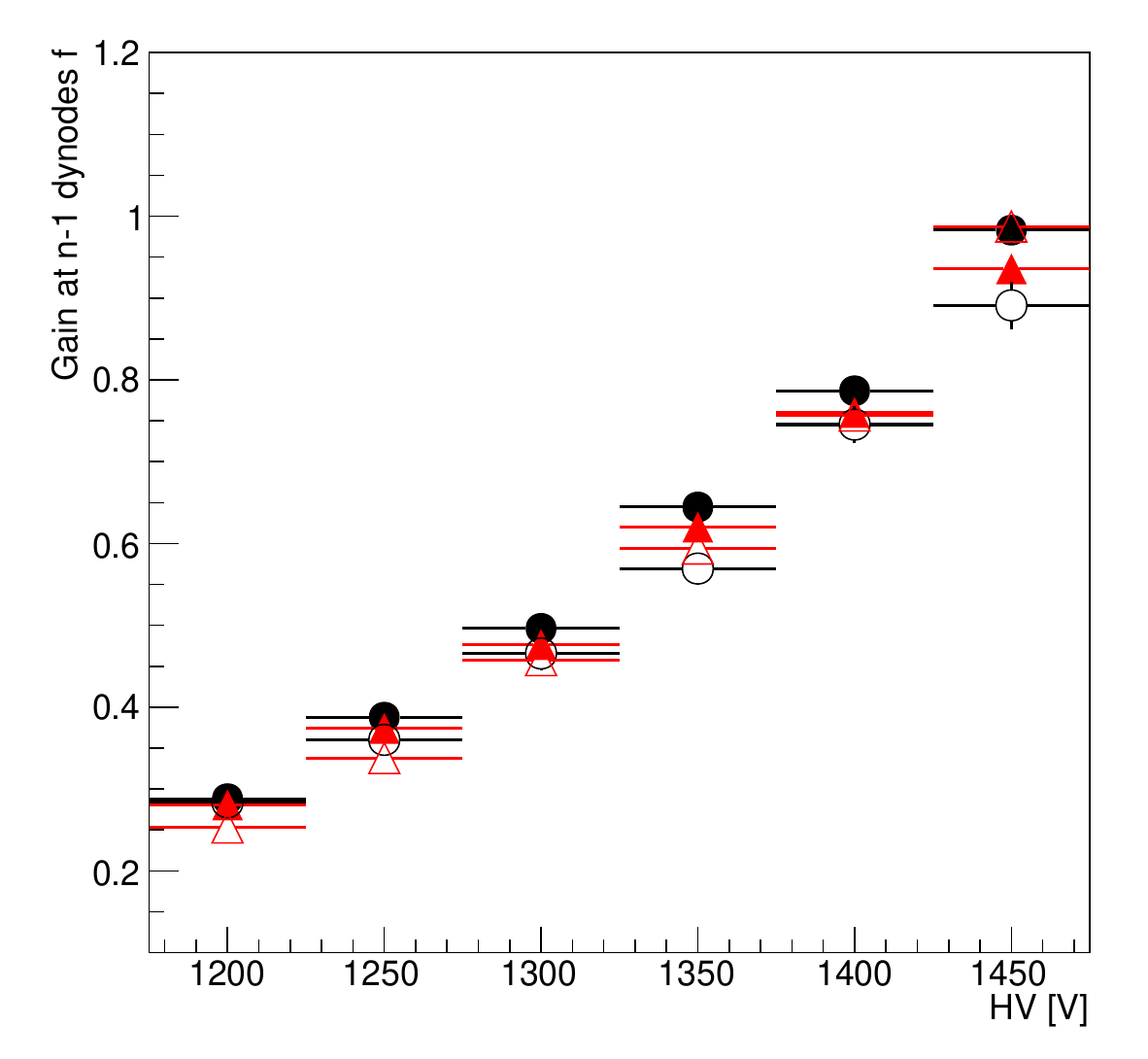}\\
    \includegraphics[width=0.45\textwidth]{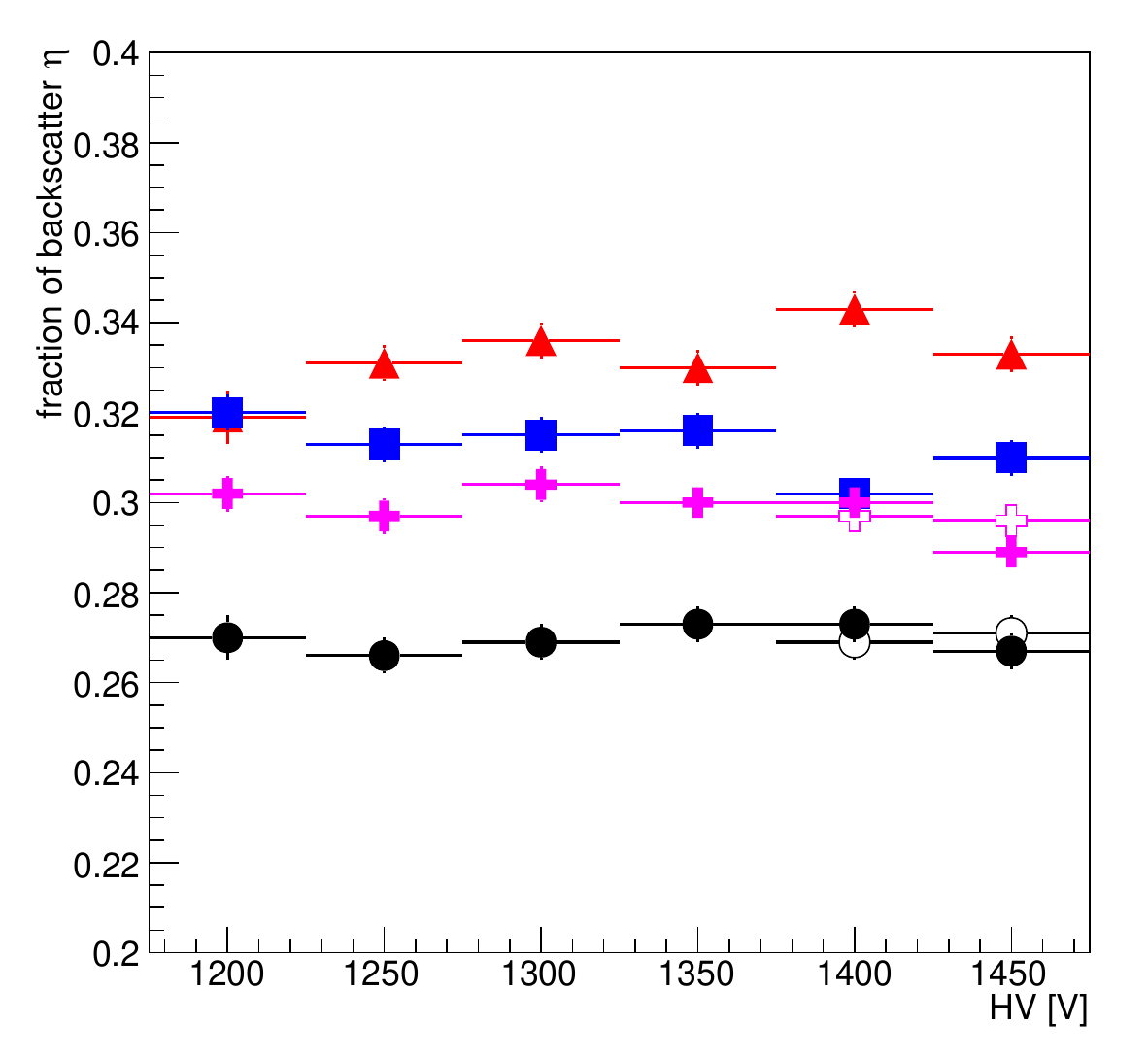}
    \includegraphics[width=0.45\textwidth]{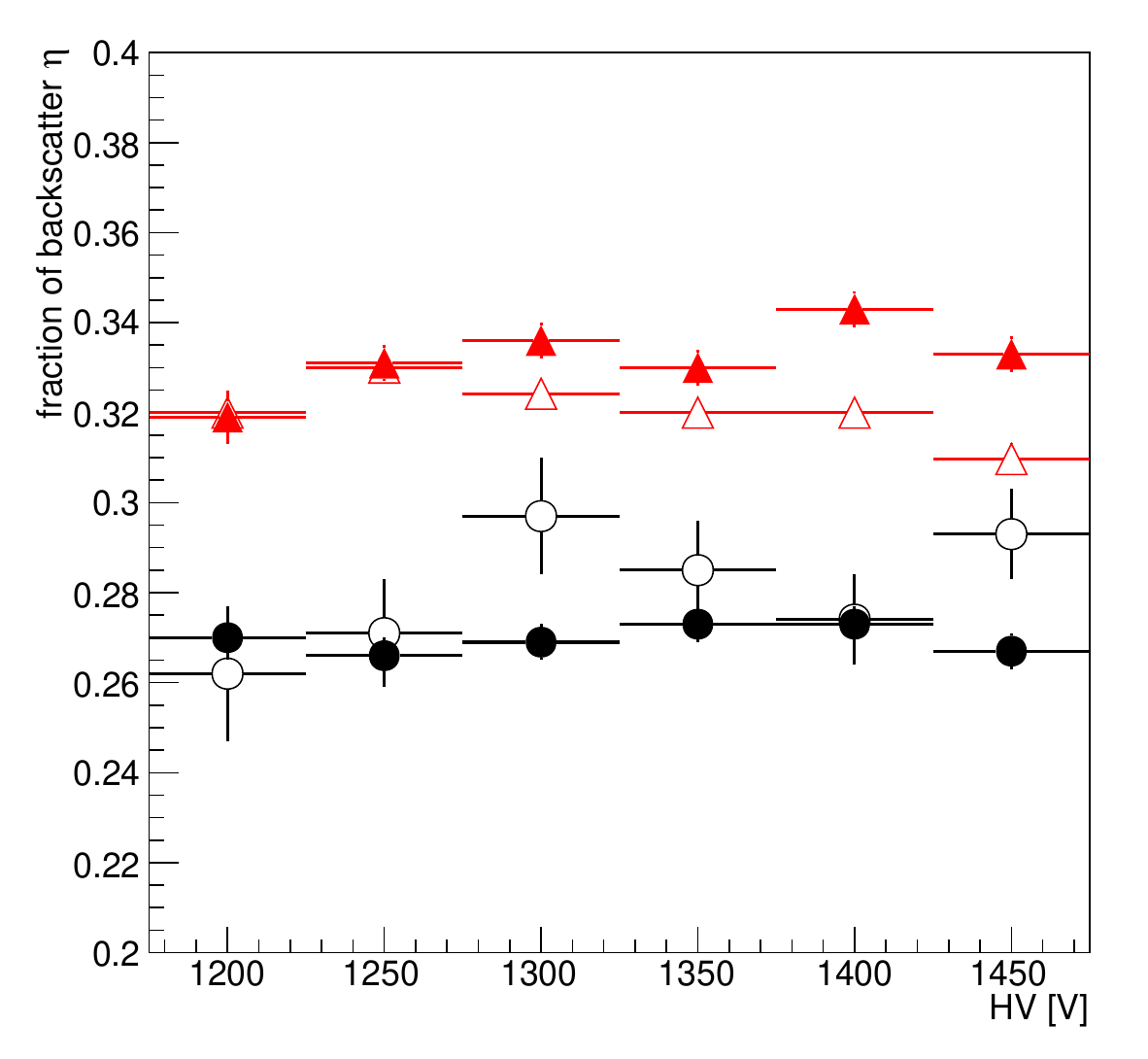}\\
    \caption{Fit parameters versus high voltage for different measurements performed (see legend). From top to bottom are shown the gain of the first dynode, $G_1$, the gain of the remaining dynodes in units of $10^6$ electrons, $f$, and the fraction of back-scattered electrons, $\eta$.}
    \label{fig:FITresultsSUMMARY1}
\end{figure*}
\begin{table*}[h!]
\caption{Fit parameters and chisquare of the model that best represent the spectral response of the Hamamatsu R5912-100 photomultiplier measured in different condition of HV, light wavelength (405~nm and 507~nm) and fibre position. In two cases (referred to as attX2) the measurements at HV=1400~V and 1450~V are repeated using a 2~dB attenuation to reduce the signal saturation observed.}\label{tab:results-R5812}
\centering
\resizebox{0.92\textwidth}{!}{
\begin{tabular}{c |  c c  c c c c | c} \hline
HV [V] & $G_{\rm 1}$  & $f\times G_{\rm 1}[\times 10^{6}]$& $R$ &   $\eta$ & $A_{\rm Exp}$ & $\alpha$ & Fit $\chi^2/{\rm ndf}$\\  \hline
\multicolumn{7}{c}{\bf Fibre in central position  - Filter 30 - $\lambda$=405~nm}\\ 
\hline

1200 &17.8$\pm$0.2 & 5.13$\pm$0.01 & 0.435 & 0.270$\pm$0.005 & 0.039$\pm$0.011 & 0.14$\pm$0.03 & 1.03 \\ 
1250 &18.0$\pm$0.2 & 6.98$\pm$0.01 & 0.435 & 0.266$\pm$0.004 & 0.042$\pm$0.006 & 0.19$\pm$0.02 & 1.11 \\ 
1300 &19.0$\pm$0.3 & 9.43$\pm$0.02 & 0.435 & 0.269$\pm$0.004 & 0.036$\pm$0.003 & 0.29$\pm$0.03 & 1.19 \\ 
1350 &19.7$\pm$0.3 & 12.70$\pm$0.02 & 0.435 & 0.273$\pm$0.004 & 0.030$\pm$0.002 & 0.39$\pm$0.04 & 1.04 \\ 
1400 &21.1$\pm$0.3 & 16.60$\pm$0.03 & 0.435 & 0.273$\pm$0.004 & 0.031$\pm$0.002 & 0.49$\pm$0.04 & 1.11 \\ 
1450 & 21.8$\pm$0.3 & 21.43$\pm$0.03 & 0.435 & 0.267$\pm$0.004 & 0.032$\pm$0.002 & 0.57$\pm$0.04 & 1.18 \\ 
1400 attX2 &20.5$\pm$0.3 & 16.90$\pm$0.03 & 0.435 & 0.269$\pm$0.004 & 0.033$\pm$0.002 & 0.45$\pm$0.03 & 1.16 \\ 
1450 attX2 &21.6$\pm$0.3 & 21.80$\pm$0.03 & 0.435 & 0.271$\pm$0.004 & 0.029$\pm$0.002 & 0.57$\pm$0.04 & 1.30 \\ 

\hline
\multicolumn{7}{c}{\bf Fibre in lateral position  - Filter 20 - $\lambda$=405~nm}\\  \hline

1200 & 18.1$\pm$0.3 & 5.08$\pm$0.01 & 0.435 & 0.319$\pm$0.006 & 0.069$\pm$0.016 & 0.14$\pm$0.02 & 1.22\\ 
1250 & 18.6$\pm$0.3 & 6.96$\pm$0.01 & 0.435 & 0.331$\pm$0.004 & 0.039$\pm$0.005 & 0.22$\pm$0.02 & 1.22\\ 

1300 &20.0$\pm$0.3 & 9.53$\pm$0.02 & 0.435 & 0.336$\pm$0.004 & 0.043$\pm$0.005 & 0.30$\pm$0.03 & 1.25 \\ 
1350 &20.3$\pm$0.3 & 12.59$\pm$0.02 & 0.435 & 0.330$\pm$0.004 & 0.041$\pm$0.003 & 0.41$\pm$0.03 & 1.15 \\ 
1400 &21.9$\pm$0.3 & 16.65$\pm$0.03& 0.435 & 0.343$\pm$0.004 & 0.035$\pm$0.002 & 0.57$\pm$0.04 & 1.14 \\ 
1450 &23.0$\pm$0.3 & 21.52$\pm$0.04 & 0.435 & 0.333$\pm$0.004 & 0.040$\pm$0.002 & 0.69$\pm$0.04 & 1.20 \\

\hline
\multicolumn{7}{c}{\bf Fibre in lateral position  at 90$^o$ - Filter 30 - $\lambda$=405~nm}\\  \hline

1200 & 18.93$\pm$0.24 & 5.161$\pm$0.008 & 0.435 & 0.320$\pm$0.004 & 0.042$\pm$0.005 & 0.15$\pm$0.02 & 1.04 \\ 
1250 &  18.94$\pm$0.25 & 7.052$\pm$0.012 & 0.435 & 0.313$\pm$0.004 & 0.049$\pm$0.009 & 0.20$\pm$0.02 & 1.24 \\ 
1300 &  20.17$\pm$0.25 & 9.596$\pm$0.015 & 0.435 & 0.315$\pm$0.004 & 0.040$\pm$0.004 & 0.30$\pm$0.03 & 1.16 \\ 
1350 &  20.99$\pm$0.26 & 12.73$\pm$0.02 & 0.435 & 0.316$\pm$0.004 & 0.039$\pm$0.002 & 0.40$\pm$0.02 & 1.02 \\ 
1400 & 21.47$\pm$0.26 & 16.75$\pm$0.02 & 0.435 & 0.302$\pm$0.003 & 0.040$\pm$0.001 & 0.50$\pm$0.03 & 1.12 \\ 
1450 &  22.32$\pm$0.30 & 21.38$\pm$0.03 & 0.435 & 0.310$\pm$0.004 & 0.041$\pm$0.001 & 0.70$\pm$0.03 & 1.07 \\

\hline
\multicolumn{7}{c}{\bf Fibre in lateral position  at 180$^o$ - Filter 30 - $\lambda$=405~nm}\\  \hline

1200 & 17.95$\pm$0.23 & 4.915$\pm$0.008 & 0.435 & 0.302$\pm$0.004 & 0.079$\pm$0.005 & 0.17$\pm$0.01 & 1.23 \\ 
1250 & 18.58$\pm$0.23 & 6.695$\pm$0.011 & 0.435 & 0.297$\pm$0.004 & 0.076$\pm$0.004 & 0.25$\pm$0.01 & 1.05 \\ 
1300 & 19.88$\pm$0.25 & 8.996$\pm$0.014 & 0.435 & 0.304$\pm$0.004 & 0.087$\pm$0.005 & 0.29$\pm$0.02 & 1.05 \\ 
1350 & 20.46$\pm$0.24 & 12.13$\pm$0.02 & 0.435 & 0.300$\pm$0.003 & 0.078$\pm$0.002 & 0.43$\pm$0.01 & 1.01 \\ 
1400 & 21.73$\pm$0.25 & 15.90$\pm$0.02 & 0.435 & 0.300$\pm$0.003 & 0.078$\pm$0.002 & 0.52$\pm$0.02 & 1.17 \\
1450 & 22.31$\pm$0.28 & 20.47$\pm$0.03 & 0.435 & 0.289$\pm$0.003 & 0.079$\pm$0.001 & 0.73$\pm$0.02 & 1.07 \\ 
1400 attX2 & 21.19$\pm$0.24 & 16.13$\pm$0.02 & 0.435 & 0.297$\pm$0.003 & 0.077$\pm$0.002 & 0.57$\pm$0.02 & 1.11 \\
1450 attX2 & 22.33$\pm$0.25 & 20.89$\pm$0.03 & 0.435 & 0.296$\pm$0.003 & 0.076$\pm$0.001 & 0.71$\pm$0.02 & 1.08  \\

\hline
\multicolumn{7}{c}{\bf Fibre in central position - Filter 40 - $\lambda$=507~nm}\\  \hline

1200 & 18.4$\pm$0.6 & 5.221$\pm$0.03 & 0.435 & 0.262$\pm$0.015 & 0.099$\pm$0.036 & 0.16$\pm$0.03 & 1.15  \\ 
1250 & 19.7$\pm$0.6 & 7.094$\pm$0.03 & 0.435 & 0.271$\pm$0.012 & 0.058$\pm$0.024 & 0.25$\pm$0.08 & 1.03 \\ 
1300 & 20.7$\pm$0.9 & \phantom{0}9.64$\pm$0.05 & 0.435  & 0.297$\pm$0.013 & 0.049$\pm$0.015 & 0.35$\pm$0.11 & 1.12 \\ 
1350 & 22.9$\pm$0.7 & 13.04$\pm$0.06 & 0.435 & 0.285$\pm$0.011 & 0.068$\pm$0.014 & 0.32$\pm$0.06 & 1.16 \\ 
1400 & 22.7$\pm$0.7 & 16.92$\pm$0.07 & 0.435 & 0.274$\pm$0.010 & 0.064$\pm$0.009 & 0.49$\pm$0.07 & 1.17  \\ 
1450 & 24.7$\pm$0.8 & 22.01$\pm$0.09 & 0.435 & 0.293$\pm$0.010 & 0.061$\pm$0.007 & 0.67$\pm$0.11 & 1.13 \\

\hline
\multicolumn{7}{c}{\bf Fibre in lateral position at 0$^o$ - Filter 30 - $\lambda$=507~nm}\\  \hline

1200 & 20.29$\pm$0.24 & 5.133$\pm$0.007 & 0.435 & 0.320$\pm$0.003 & 0.0510$\pm$0.004 & 0.145$\pm$0.010 & 1.12 \\ 
1250 &  20.99$\pm$0.24 & 7.091$\pm$0.010 & 0.435 & 0.330$\pm$0.003 & 0.0408$\pm$0.003 & 0.227$\pm$0.015 & 1.32 \\ 
1300 &  20.97$\pm$0.24 & 9.595$\pm$0.013 & 0.435 & 0.324$\pm$0.003 & 0.0470$\pm$0.0028 & 0.29$\pm$0.02 & 1.29 \\ 
1350 &  21.89$\pm$0.26 & 13.00$\pm$0.02 & 0.435 & 0.320$\pm$0.003 & 0.0451$\pm$0.0017 & 0.39$\pm$0.02 & 1.16 \\ 
1400 & 22.57$\pm$0.28 & 17.07$\pm$0.02 & 0.435 & 0.320$\pm$0.003 & 0.0442$\pm$0.0013 & 0.51$\pm$0.02 & 1.12 \\ 
1450 &  22.37$\pm$0.34 & 22.09$\pm$0.04 & 0.435 & 0.310$\pm$0.004 & 0.0448$\pm$0.0010 & 0.72$\pm$0.03 & 1.36  \\
\hline
\end{tabular}
}
\end{table*}
\begin{figure*}[!h]
    \includegraphics[width=0.45\textwidth]{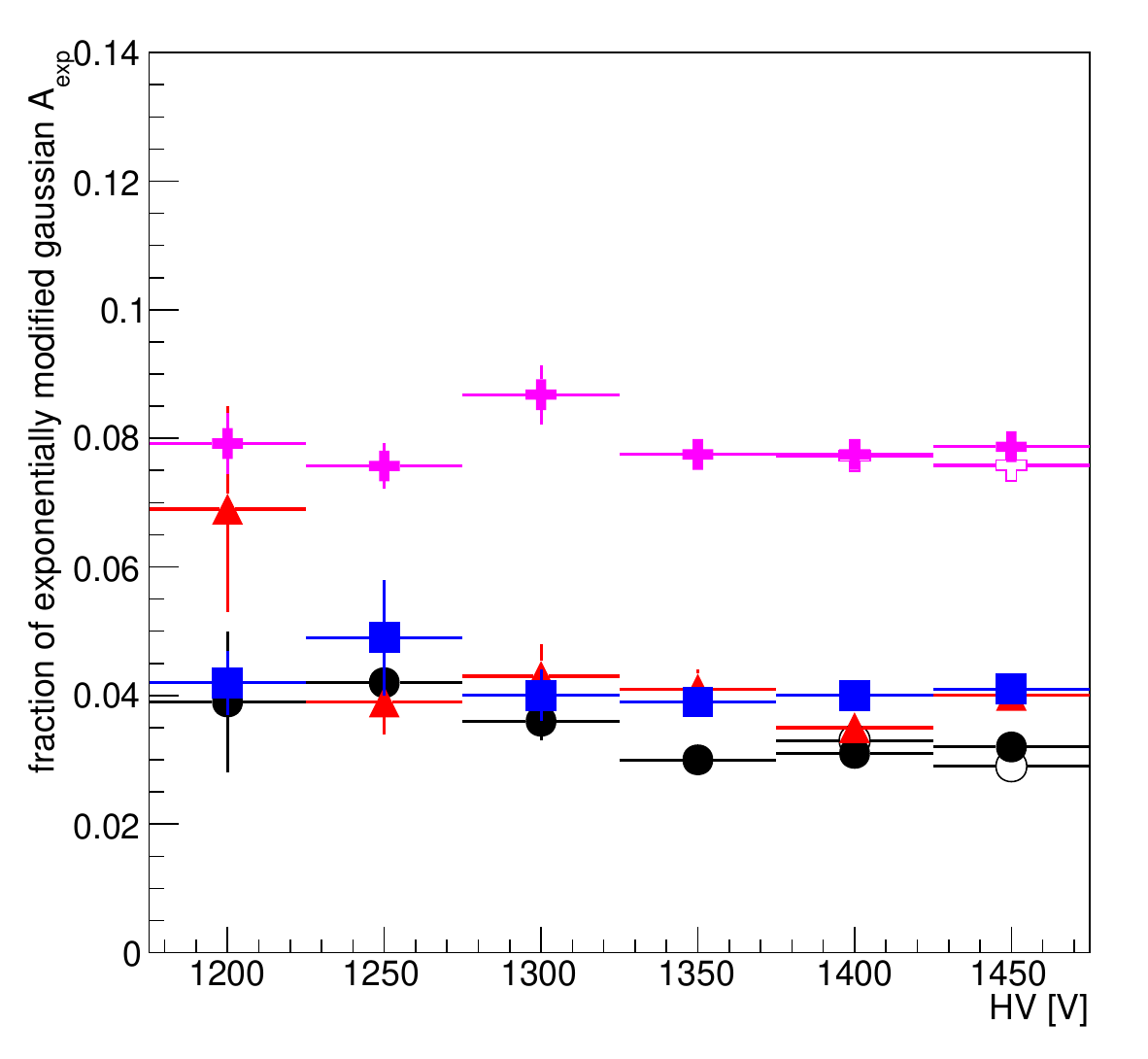}
    \includegraphics[width=0.45\textwidth]{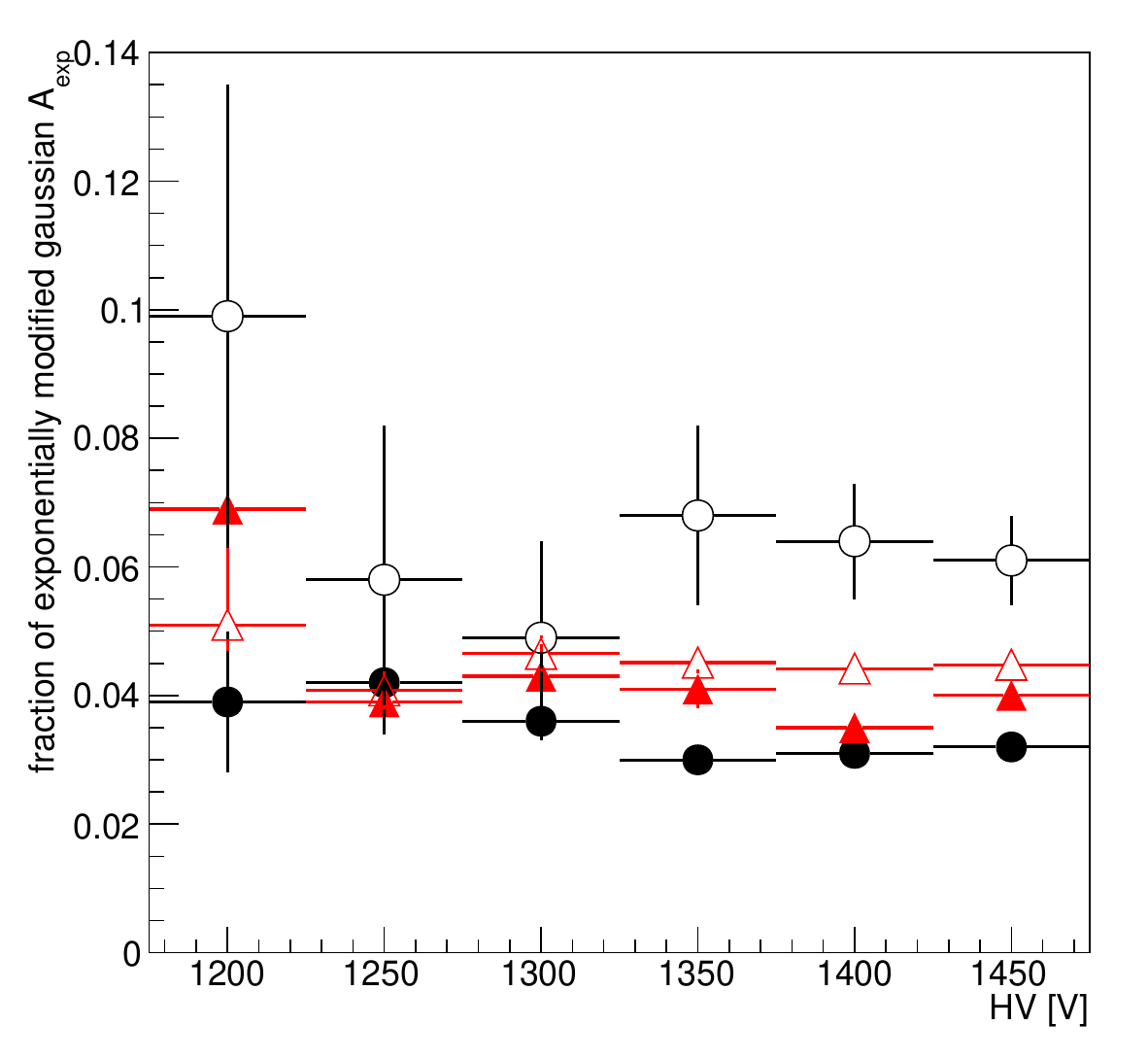}\\
    \includegraphics[width=0.45\textwidth]{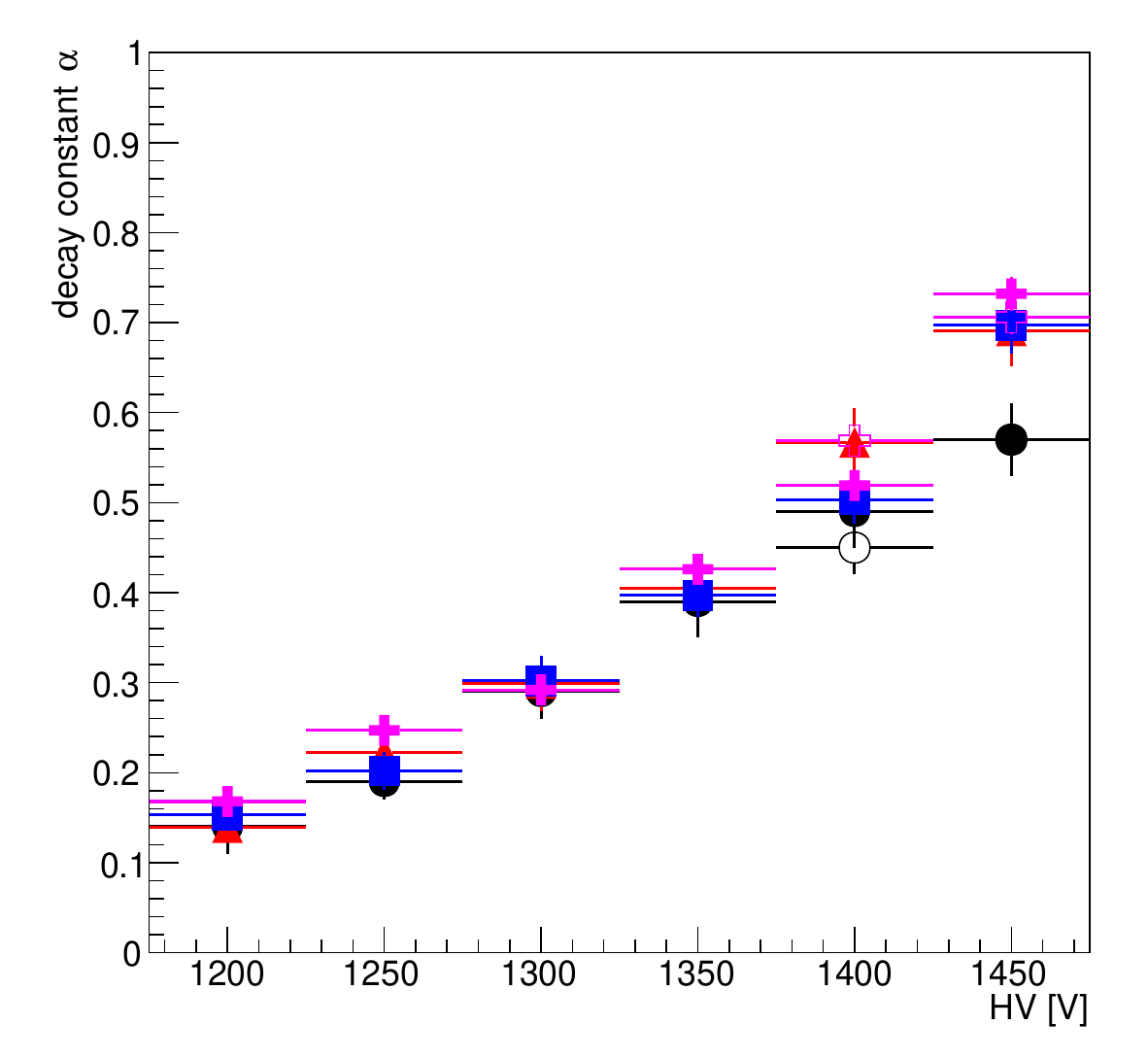}
    \includegraphics[width=0.45\textwidth]{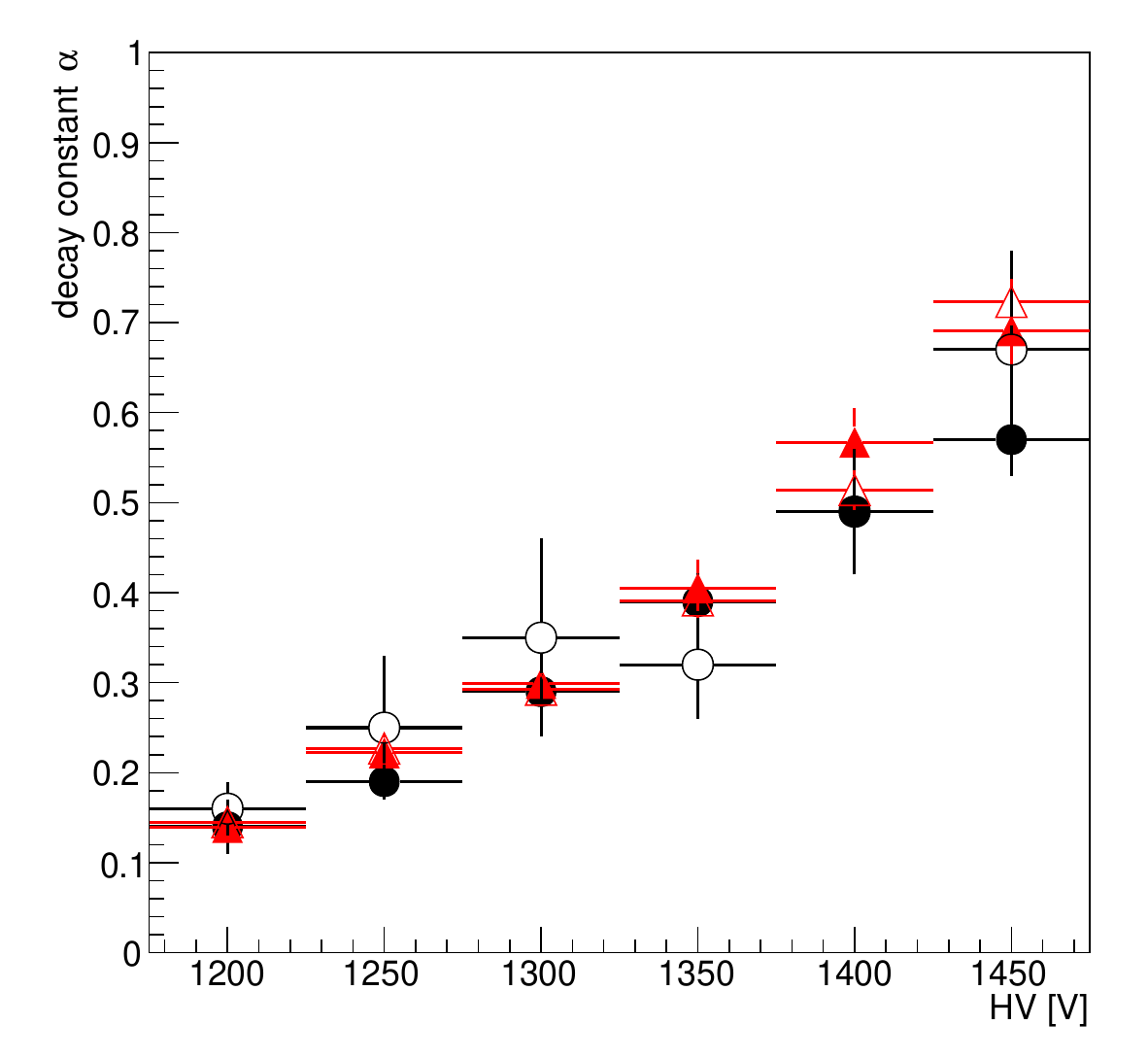}\\
    \caption{Fit parameters versus high voltage for different measurements performed. From top to bottom are shown the fraction of exponentially modified gaussian, $A_{\rm Exp}$, and its decay constant, $\alpha$. The figure legend is the same as in Fig.~\ref{fig:FITresultsSUMMARY1}. }
    \label{fig:FITresultsSUMMARY2}
\end{figure*}

\subsection{Photomultiplier Hamamatsu R5912-100}
Figure~\ref{fig:my_label} shows the fit results for one of the acquired data sample. 
The fit is performed for charge values larger than $0.3\times10^{6}$ electrons, where the pedestal subtraction is reliable, as proved by using alternative subtraction methods.
The fit nicely represents the whole spectrum as demonstrated by the normalized chisquare value, $\chi^2/{\rm ndf} = 1.04$, and by the normalized residuals (\emph{pulls}) also shown in the figure, which show a flat distribution around zero. 

The fit requires the contributions of one and two photoelectrons ($1-A_{\rm 2PE} \sim94\%$, and $A_{\rm 2PE} \sim 6$\%), while the contribution of three photoelectrons is found to be negligible, so it is set to zero. The SPE model is found to be described by a significant ($\eta\sim 27$\%) contribution due to back-scattered photoelectrons and a smaller, but significant ($A_{\rm Exp}\sim 4$\%), exponential term due to low-charge events. The pre-pulse contribution instead is found to be negligible, so it is set to zero.

The fit to the spectra of the other data samples acquired have a quality level similar to the one of Figure~\ref{fig:my_label}, with $\chi^2/{\rm ndf}$ values in a range $[1.03, 1.3]$ demonstrating the validity of the model developed in a wide range of conditions. They are listed in Table~\ref{tab:results-R5812} together with the numerical results of the fit parameters.

Figures~\ref{fig:FITresultsSUMMARY1} and~\ref{fig:FITresultsSUMMARY2} show the dependence of the fit parameters on the HV for the different measurements taken with the fibre at different positions and at different wavelengths.
As expected the gain at the first dynode $G_1$ increases approximately linearly with the HV with the same percentage increase. 
The gain at the remaining dynodes $f$ also increases with the HV, but with a larger percentage increase, consistent with the number (nine) of remaining dynodes contributing to the gain.
Also the parameter $\alpha$ shows a linear dependency on the HV. 
For these three parameters, there does not seem to be a dependence on the wavelength or the position of the fibre. 
On the contrary, the fraction of partially amplified electrons, $\eta$, and of the very low charge events, $A_{\rm Exp}$, do not depend on the HV (within the range of values considered) but do depend on the position of the fibre.

\subsection{Photomultiplier Hamamatsu 6233}
\begin{figure}[h!]
    \centering
    \includegraphics[width=0.46\textwidth]{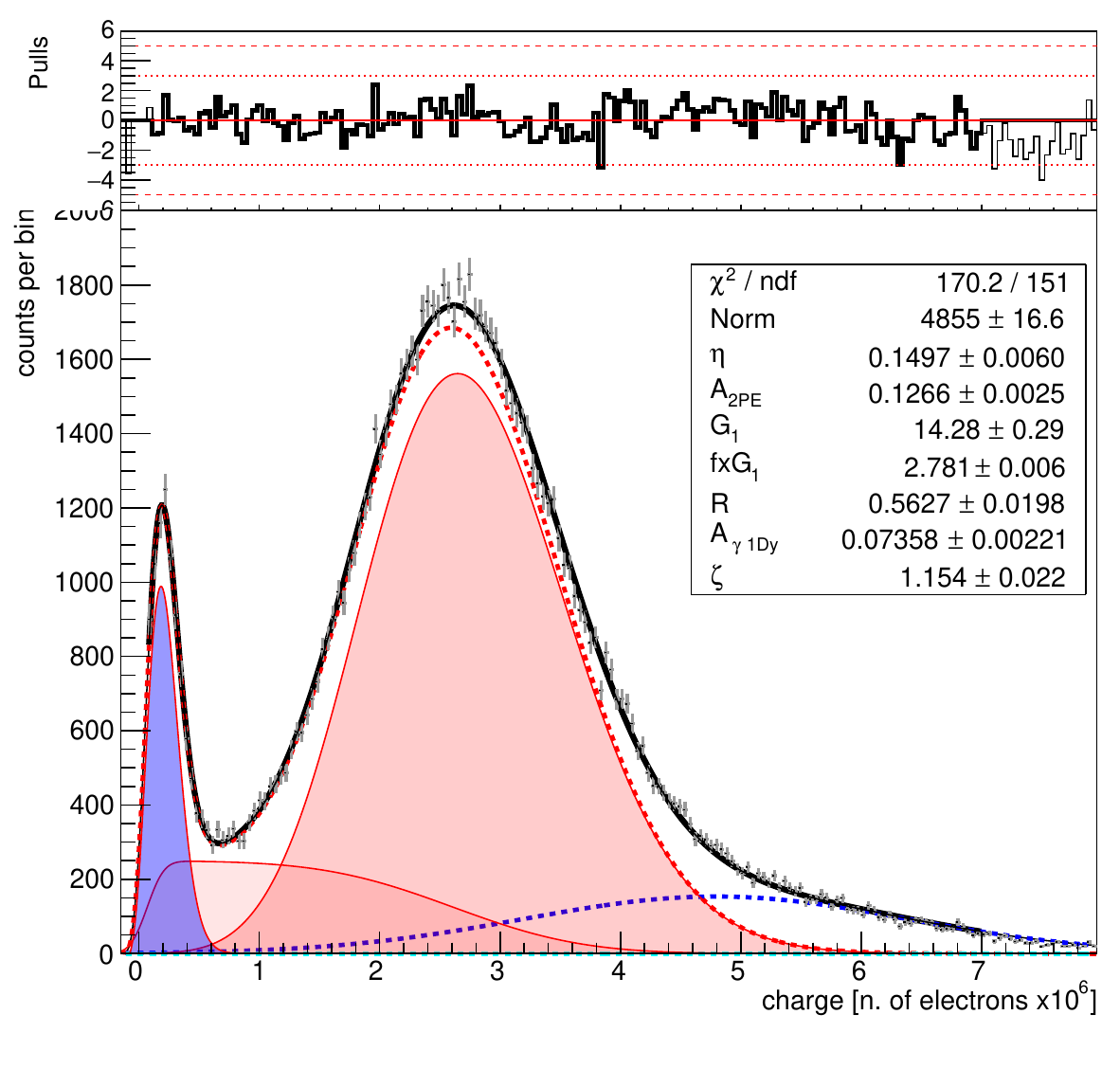}\\
\caption{Fit to the spectral response to light-only for the photomultiplier Hamamatsu 6233 with bias voltage HV=1300~V, illuminated with light with a wavelength of 405~nm and fibre in central position.
The fit is performed in the interval $[0.08,7]\times 10^6 \,e$ and the result is shown by the thick black line. 
The distribution of the normalized residuals (\emph{pulls}) is shown in the top panel.
}
    \label{fig:Fit_Hamamatsu6233}
\end{figure}

Figure~\ref{fig:Fit_Hamamatsu6233} shows the fit to one of the acquired data sample of photomultiplier Hamamatsu 6233.
The fit range begins at a charge value of $0.08\times10^{6}$ electrons, benefiting from the lower electronic noise and narrower pedestal compared to photomultiplier R5912-100. An upper value of the fit range is also set to exclude the region of the spectrum where the contribution of two photoelectrons dominates and its modelling with the Gaussian shows some deviations; typically it is set to $\sim 2\mu $ (see Eq.~\eqref{eq:cont_poisson1}).

\begin{table*}[h!]
\caption{Fit parameters and chisquare of the model that best represent the spectral response of the Hamamatsu 6233 photomultiplier measured in different conditions. }\label{tab:results-6233}
\centering
\resizebox{1\textwidth}{!}{
\begin{tabular}{c |  c c c c  c c  | c } \hline
HV [V] & $G_{\rm 1}$ & $f \times G_1$ [$\times 10^{6}$] & $R$ &  $\eta$ & A$_{ \gamma\rm 1Dy}$  & $\zeta$ & Fit $\chi^2/{\rm ndf}$\\  \hline
\multicolumn{8}{c}{\bf Fibre in central position  - Filter 33 - $\lambda$=405~nm}\\ 
\hline
1300 &14.3$\pm$0.3 & 2.78$\pm$0.006 & 0.56$\pm$0.02 & 0.150$\pm$0.006 & 0.074$\pm$0.002 & 1.16$\pm$0.02 & 1.12  \\
1350 & 14.7$\pm$0.3 & 3.44$\pm$0.007 & 0.53$\pm$0.02 & 0.170$\pm$0.006 & 0.073$\pm$0.002 & 1.17$\pm$0.02 & 1.10  \\
1400 & 14.9$\pm$0.3 & 4.17$\pm$0.009 & 0.53$\pm$0.02 & 0.174$\pm$0.006 & 0.073$\pm$0.002 & 1.15$\pm$0.02 & 1.04  \\ 
1450 & 15.2$\pm$0.3 & 5.03$\pm$0.011 & 0.55$\pm$0.02 & 0.171$\pm$0.006 & 0.077$\pm$0.002 & 1.16$\pm$0.02 & 1.12  \\ 
\hline
\end{tabular}
}
\end{table*}

The fit requires the contributions of one and two photoelectrons ($1-A_{\rm 2PE} \sim87\%$, and $A_{\rm 2PE} \sim 13$\%), while the contribution of three photoelectrons is found to be negligible, so it is set to zero. Compared to the photomultiplier R5912-100, a smaller contribution due to back-scattered photoelectrons ($\eta\sim 17$\%) and a significant pre-pulse contribution ($A_{\gamma 1\rm Dy} \sim 7$\%) are found, while no significant exponential term due to very-low charge events is needed.

The fit results for the data samples acquired at different bias voltage are listed in Table~\ref{tab:results-6233}. Also in this case the parameters $G_1$ and $f$ depend on the HV, while the fractions $\eta$ and $A_{\gamma 1{\rm Dy}}$ are constant, within the uncertainties. 
As mentioned before, the parameter $R$ is determined by the fit and found to be consistent with the estimated value of 0.5 from simulation. Similarly the parameter $\zeta$ is found to be larger than and close to one, as expected (see Eq.~\eqref{eq:gian_pre-pulse}).

\section{Conclusions}
In this work we have developed an analytical model for the single photoelectron (SPE) response of single-photon-counting photomultipliers, with particular emphasis on the charge region between the electronics pe\-destal and the fully amplified photoelectron peak.

The model is based on the physical description of back-scattering of primary photoelectrons at the first dynode and provides an explicit parametrization of partially amplified photoelectrons. Analytical descriptions of the fully amplified peak and of very low-charge signals are also derived within a consistent framework.
The model was implemented to fit the SPE response of two different models of photomultipliers, Hamamatsu R5912-100 and Hamamatsu 6233, using a dedicated setup with a low-noise amplifier, to study in detail the region at lower collected charge.

The model was tested against a wide set of measurements acquired under different operating conditions, including variations of the applied high voltage, illumination position, and wavelength. In all cases, the model provides a good description of the measured charge spectra, with fit qualities close to unity. The extracted gain parameters exhibit the expected dependence on the applied high voltage, while the fractions of partially amplified and low-charge events are found to be largely independent of the bias voltage within the explored range.

Overall, the proposed approach offers a physically motivated and robust description of the SPE response of photomultipliers, avoiding the use of empirical para\-metrizations for the intermediate-charge region. Owing to its limited number of intrinsic parameters and its validity across different photomultiplier types and operating conditions, the model is well suited for use in calibration and simulation frameworks of experiments requiring precise single-photon characterization.

\section*{Acknowledgments}

We acknowledge support from the Istituto Nazionale di Fisica Nucleare (Italy) and Laboratori Nazionali del Gran Sasso (Italy) of INFN. 
We also acknowledge the valuable discussions on the model that took place within the XENONnT Neutron Veto group.

\appendix
\section{Detailed formulation}
\subsection{The scaled continuous Poisson distribution}
\label{sec:appendix1}
In this section we derive the probability distribution for fully amplified photoelectrons hitting the first dynode as a function of typical parameters related to photomultipliers. Assuming that the number of electrons emitted at the first dynode follows a Poisson statistics, the mean and the variance of the distribution are given by $G_1$ and $\sigma_1^2 = G_1$, respectively, where $G_1$ is the gain at the first dynode.
For each of the following dynodes $i$ ($2 \le i \le N$) we don't distinguish between fully amplified and partially amplified electrons and  therefore we refer to the mean gain $\langle G_i\rangle$ and to the corresponding variance $\sigma_i^2$. The resulting total gain $g$ of the photomultiplier  is then 
\begin{align}
    g = G_1\cdot \langle G_2\rangle \cdot...\cdot \langle G_N\rangle \, ,
\end{align}
and the variance, $\sigma_g^2$, of the distribution of the number of electrons resulting from the cascade process along the $N$ dynodes is

\begin{eqnarray}
\begin{array}{l} 
    \sigma_g^2 =g^2\left(\frac{\sigma_1^2}{G_1^2}+\frac{\sigma_2^2}{G_1\cdot \langle G_2\rangle ^2}+...+\frac{\sigma_N^2}{G_1\cdot \langle G_2\rangle \cdot ...\cdot \langle G_N\rangle ^2}\right)\\ 
    =\frac{g^2}{G_1}\left(1+\frac{\sigma_2^2}{\langle G_2\rangle ^2}+\frac{\sigma_3^2}{\langle G_2\rangle \cdot \langle G_3\rangle ^2}+...+\frac{\sigma_N^2}{\langle G_2\rangle \cdot ...\cdot \langle G_N\rangle ^2}\right) \\
    =\frac{g^2}{G_1}\left(1+R^2\right) ,
\end{array}
\end{eqnarray}
where one should notice that the term $R^2$ represents the variance divided  by the square of the average gain of the remaining $N-1$ dynodes, $R^2 = {\sigma^2_{N-1}}{/g_{N-1}^2}$,\footnote{For completeness the complete formula is the following
\begin{footnotesize}
\begin{align}
\nonumber
R^2 = \left(\frac{\sigma_2^2}{\langle G_2\rangle^2}+\frac{\sigma_3^2}{\langle G_2\rangle\cdot \langle G_3\rangle^2}+...+\frac{\sigma_N^2}{\langle G_2\rangle\cdot \langle G_3\rangle\cdot ...\cdot \langle G_N\rangle^2}\right) 
\end{align}
\end{footnotesize}
} being $g_{N-1}=\langle G_2\rangle\cdot \langle G_3\rangle\cdot...\cdot \langle G_N\rangle$.
According to the central limit theorem, which holds for \mbox{$n\gtrsim 10$}, if $n$ secondary electrons are emitted from the first dynode, the response of the $N-1$ successive dynodes will approximate a Gaussian distribution with a resolution $R/\sqrt{n}$.


Let us now introduce a conversion factor, which converts the gain of the $N-1$ dynodes $g_{N-1}$ into ADC units, $f$, such that the total  photomultiplier gain in ADC units is given by $\mu =  G_1 f$, and the resolution of the $N-1$ dynodes, also in ADC units, is $\Sigma = Rf$. 
Additionally, we include the electronic readout noise in ADC units, 
$\sigma_{\rm ped}$, whose contribution is added after that due to the multiplication chain.
The resulting variance $\sigma^2$, in ADC units, for  fully amplified primary photoelectrons is therefore (compare to \eqref{eq:sigma_FA}):
\begin{align}
\label{eq:varADC}
    \sigma_{{\mathscr P}_{\rm FA}}^2 = \sigma^2 = \sigma^2_g +\sigma_{\rm ped}^2 =  G_1f^2\left(1+R^2\right)+\sigma_{\rm ped}^2.
\end{align}
Assuming that the final probability density has a continuous poissonian shape of the form~\cite{Ilienko, MARSAGLIA19861187}
\begin{align}
{\mathscr P}_{\rm FA}(x) = \rho\frac{e^{-\rho\mu}\left(\rho\mu\right)^{\rho x}}{\Gamma(1+\rho x)},
\label{eq:cont_poisson}
\end{align}
with a scaling factor $\rho = \mu/\sigma^2_{{\mathscr P}_{\rm FA}}$ that is implied by the properties of a scaled Poisson distribution, one finds
\begin{align}
\rho = \frac{G_1f}{G_1f^2\left(1+R^2\right)+\sigma_{\rm ped}^2}\,.
\end{align}
\begin{figure}[t]
    \centering
    \includegraphics[width=0.6\columnwidth]{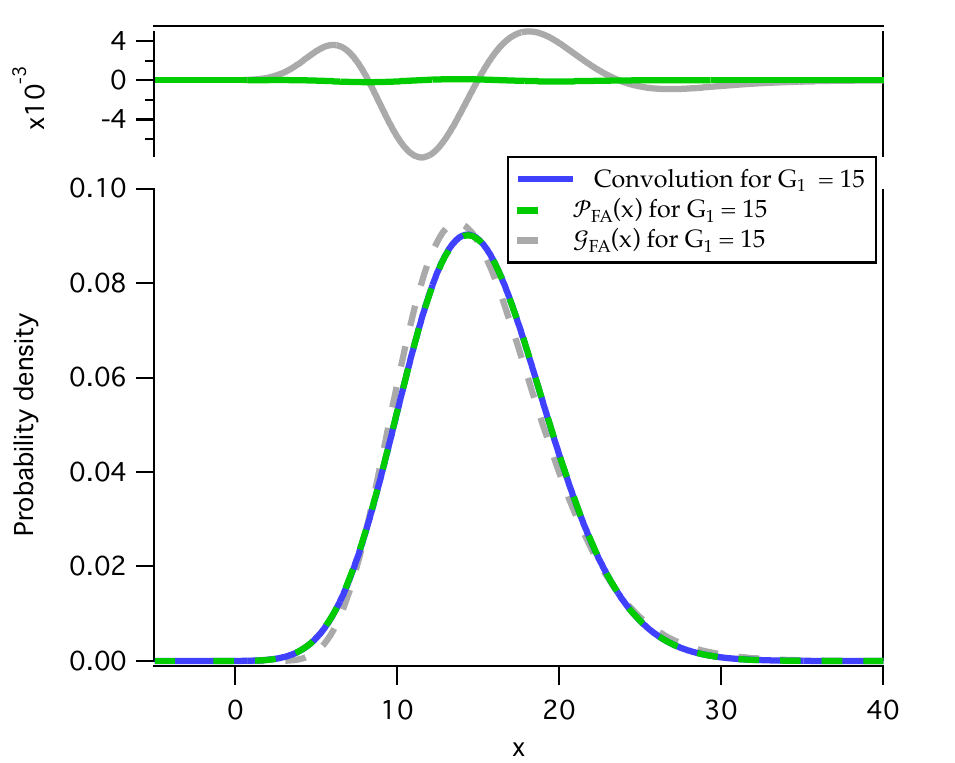}
    \caption{Comparison of the probability density function ${\mathscr P}_{\rm FA}(x)$ of Eq.~\eqref{eq:cont_poisson} with the direct convolution of the discrete Poisson distribution in Eq.~\eqref{eq:Poisson} with the Gaussian function in Eq.~\eqref{eq:gauss}. Here we have set $f = 1$, $G_1 = 15$, $R = 0.5$ and $\sigma_{\rm ped}/f = 1$.}
    \label{fig:Poisson}
\end{figure}
This function was verified numerically for $R^2G_1 \gtrsim 1$ 
by convolving the Poisson distribution \eqref{eq:Poisson} with a Gaussian  with standard deviation in ADC units
$\sigma_{\rm eff}(n)$ depending on the number $n$ of secondary electrons generated at the first dynode with gain $G_1$:
\begin{align}
\label{eq:gauss}
    \begin{split}
    \sigma_{\rm eff}(n)=&\sqrt{n\Sigma^2+\sigma_{\rm ped}^2}\; ,\\
    G(x,n) =& \frac{1}{\sqrt{2\pi}\sigma_{\rm eff}}e^{-\frac{1}{2}\left(\frac{x-n}{\sigma_{\rm eff}}\right)^2}.
    \end{split}
\end{align}

Figure~\ref{fig:Poisson} shows a comparison of three curves: the dashed green curve following Eq. \eqref{eq:cont_poisson} and the continuous blue curve following the convolution of Eq.~\eqref{eq:Poisson} with Eq.~\eqref{eq:gauss}. The grey dashed curve is described below. The parameters $G_1 = 15$, $R=0.5$ and $\sigma_{\rm ped}/f=1$ are fixed. The difference between the first two curves is below $2\times 10^{-4}$ (the residuals are shown in green) proving the validity of the assumption in Eq. \eqref{eq:cont_poisson}.  By construction, the mean and variance of ${\mathscr P}_{\rm FA}(x)$ are respectively:
\begin{align}
\begin{split}
\langle {\mathscr P}_{\rm FA}\rangle  & = \mu= G_1f ,\\
\sigma^2_{{\mathscr P}_{\rm FA}} &=  \frac{\mu}{\rho} = G_1f^2\left(1+R^2\right)+\sigma^2_{\rm ped}.
\end{split}
\end{align}
Often the Gamma distribution
\begin{equation}
\label{eq:Gamma}
    {\mathscr G}_{\rm FA}(x) = \frac{\beta}{\mu}\frac{\left(x\frac{\beta}{\mu}\right)^{\beta-1}e^{-x\frac{\beta}{\mu}}}{\Gamma(\beta)}
\end{equation}
is used to describe the fully amplified component of the SPE.
The mean and variance of the Gamma distribution are respectively $\mu$ and 
 $\mu^2/\beta$ and must result in:
\begin{align}
\begin{split}
\langle {\mathscr G}_{\rm FA}\rangle  & = \mu = G_1f, \\
\sigma^2_{{\mathscr G}_{\rm FA}} &=  \frac{\mu^2}{\beta} = G_1f^2\left(1+R^2\right)+\sigma^2_{\rm ped}.
\end{split}
\end{align} 
In Figure \ref{fig:Poisson} we have superimposed the Gamma distribution, grey dashed curve, considering the parameters $G_1=15$, $R=0.5$ and $\sigma_{\rm ped}/f=1$. The residuals in this case are shown in grey. 
Clearly the scaled continuous Poisson distribution is far superior to the Gamma distribution in describing the continuous blue curve. Note that even leaving the parameters $\mu$ and $\beta$ free, the residuals decrease only by a factor 2. This justifies the use of \eqref{eq:cont_poisson} to describe the fully amplified single photoelectron signals.

\subsection{Partially amplified electrons}\label{sec:appendix2}
In this section we derive the continuous probability density distribution for partially amplified photoelectrons that release only part of their energy on the first dynode due to back-scattering. 

\begin{figure}[b]
    \centering
    \includegraphics[width=0.7\columnwidth]{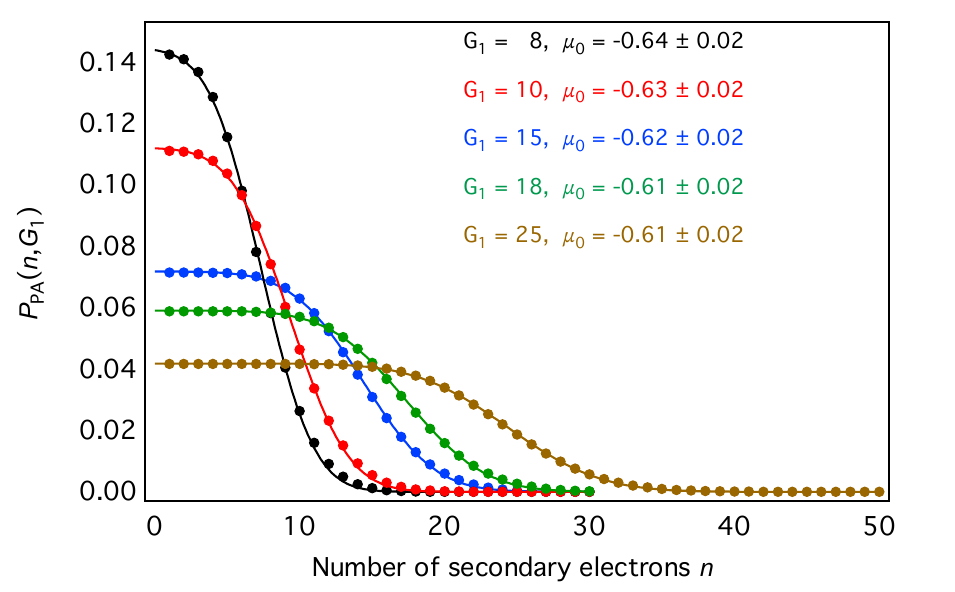}
    \caption{Dots: Plots of the discrete function given in Eq.~\eqref{eq:cassetta} for various values of the gain $G_1$. Lines: Fitting results obtained with Eq.~\eqref{eq:cassetta_norm_bis} leaving the parameter $\mu_0$ free. An average value $\mu_0 = -0.62$ is found.}
    \label{fig:mu0}
\end{figure}

The discrete distribution of \emph{detectable} secondary electrons ($n>0$) emitted by the first dynode in case of back-scattered photoelectrons follows the discrete probability density function 
\begin{align}
    P_{\rm PA} (n,G_1) = {\cal P} (n,G_1)/F  \quad \text{for} \quad n>0 \, ,
\end{align}
given in Eqs.~\eqref{eq:cassetta} and~\eqref{eq:cassetta_norm}, and reported here to simplify reading.
This equation can be approximated by a smoothed step function as
\begin{equation}
\label{eq:cassetta_norm_bis}
    P_{\rm PA} (n,G_1) \propto 
    \left[1-{\rm erf}\left(\frac{n-G_1+\mu_0}{\sqrt{2G_1}}\right)\right]\,\text{for}\, n>0\, , 
\end{equation}
where $G_1$ is the gain at the first dynode and, as we will see, $\mu_0$ is a constant independent of $G_1$.
The independence of $\mu_0$ from the gain $G_1$ was determined by fitting 
 Eq. \eqref{eq:cassetta_norm} with Eq. \eqref{eq:cassetta_norm_bis} leaving $\mu_0$ as a free parameter for different values of $G_1$. The results of various fits for the gain in the range $8\le G_1\le 25$ is shown in Figure \ref{fig:mu0}. The average value $\mu_0 = -0.62$ was chosen.

 The successive amplification by the remaining $N-1$ dynodes introduces a smoothing of the distribution for $n>0$, which is modelled by convolving Eq.~\eqref{eq:cassetta_norm_bis} with the Gaussian function Eq.~\eqref{eq:gauss}. Given the asymmetric rounded `box' shape of the resulting distribution we described the convolution with the normalized function as:
\begin{align}
\label{eq:cassetta_conv_bis}
    {\mathscr P}_{\rm PA}(x) = \frac{\left[1+{\rm erf}\left(\frac{x-\mu_L}{\sqrt{2}\sigma_L}\right)\right]\left[1-{\rm erf}\left(\frac{x-\mu_R}{\sqrt{2}\sigma_R}\right)\right]}{4(\mu_R-\mu_L)}\, .
\end{align} 
The main parameters are $\mu_L$ and $\mu_R$, describing the position of the rise and fall at the edges, and $\sigma_L$ and $\sigma_R$ describing the rise and fall at the edges. By construction and for $n \ge 1$ one has $\mu_R > \mu_L$ and therefore the condition $\mu_R = \mu_L$ is never verified.

Note that the convolution of the discrete step function $P_{\rm PA} (n,G_1)$ at $n=1$ [see Eq.~\eqref{eq:cassetta_norm_bis}] with a Gaussian function whose standard deviation is {\it constant} will leave the flex of the resulting continuous distribution at $x_{\rm flex} = 0.5$. In the case considered here, though, the smoothing Gaussian function [Eq.~\eqref{eq:gauss}] has a standard deviation which depends on  $\sqrt{n}$  which shifts the flex position from the value $x_{\rm flex}=0.5$ to lower values. 
A study of this shift was performed by fitting the result of the convolution of $P_{\rm PA} (n,G_1)$ with Eq.~\eqref{eq:cassetta_conv_bis}
considering different gains at the first dynode between $5\le G_1 \le 25$ and a range of the electronic noise $0.05\le\sigma_{\rm ped}/f\le 0.85$ as a function of the resolution $R$ in the range $0\leq R \leq 1$.

\begin{figure}[t]
    \centering
    \includegraphics[width=0.7\columnwidth]{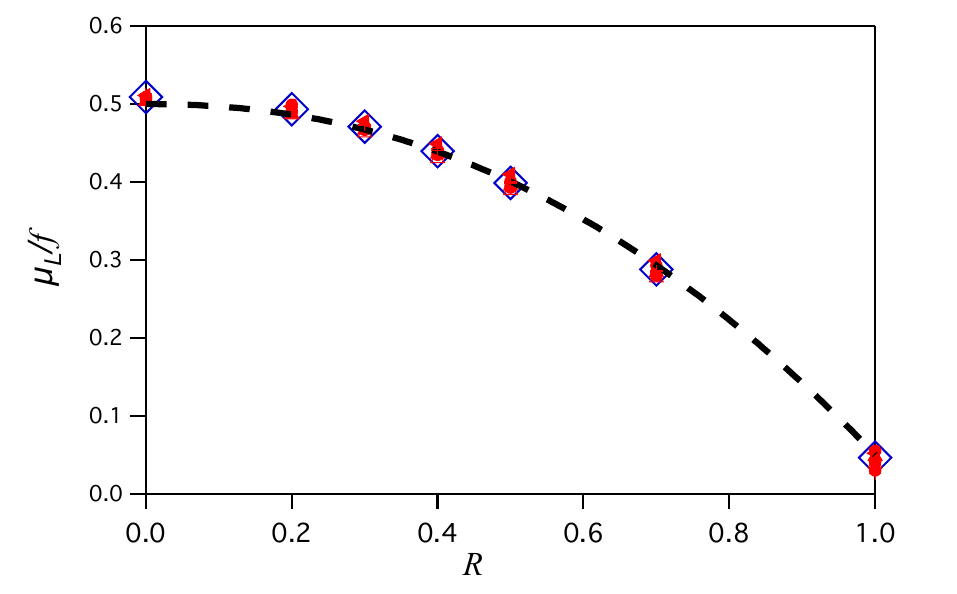}
    \caption{Dependence of $\mu_L/f$ as a function of $R$ for different combinations of $G_1$ and $\sigma_{\rm ped}/f$ (red dots). The dashed black line represents the best fit (see text).}
    \label{fig:m1_vs_sped}
\end{figure}

Interestingly the value of 
$\mu_L/f$ only depends on $R$ as can be seen in Figure~\ref{fig:m1_vs_sped}. The superimposed dashed curve is a fit with a power function, with the 
$y$ intercept fixed at $x_{\rm flex}=0.5$, resulting in 
\begin{align}
    \mu_L = f\left(0.5-0.45R^{2.2}\right).
\end{align}
\begin{figure}[t]
    \centering
    \includegraphics[width=0.7\columnwidth]{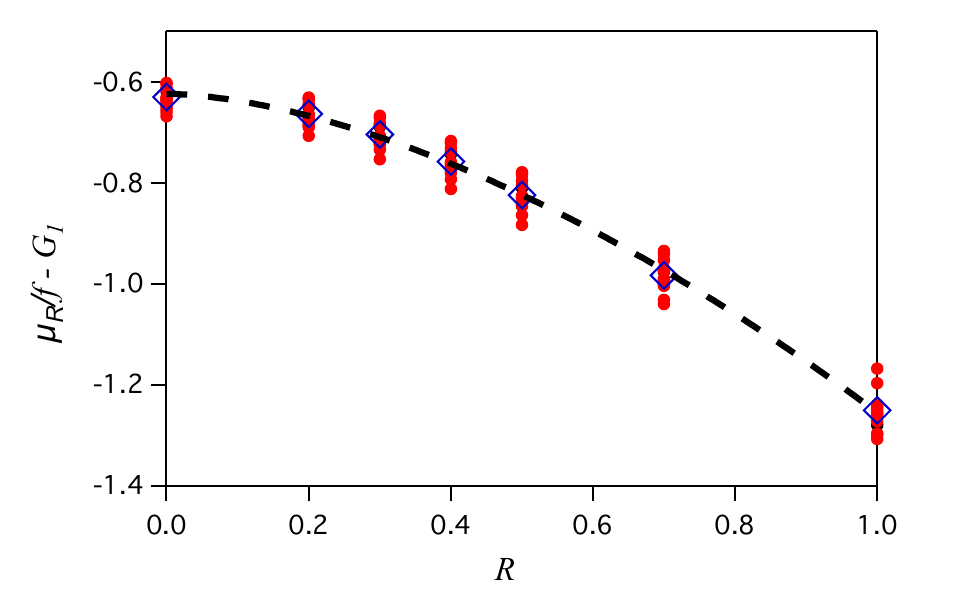}
    \caption{Dependence of $\mu_R/f - G_1$ as a function of $R$ for different combinations of $G_1$, $\sigma_{\rm ped}$ and $R$ (red dots). The dashed black line represents the best fit (see text).}
    \label{fig:A_vs_sped}
\end{figure}
Here we have used the central values of the fit parameters for the power function whose relative errors are $\lesssim 2\%$.
For $\mu_R/f$, shown in Figure~\ref{fig:A_vs_sped}, one finds a dispersion of $\Delta \mu_R/f \approx 0.05$ for each value of $R$ depending on $G_1$. Again we determined a unique function independent of the gain $G_1$ and electronic noise $\sigma_{\rm ped}/f$. 
The $\mu_R/f$ values were fitted with the function
\begin{align}
    \mu_R = f\left(G_1+\mu_0-AR^{\gamma}\right),
\end{align} 
with best fit parameters $\mu_0 = -0.62\pm0.04$, $A = 0.63\pm0.06$ and $\gamma = 1.7\pm0.3$. Note that the value of $\mu_0$ is in agreement with  the value reported for Eq. \eqref{eq:cassetta_norm_bis}, as it should be, resulting from Figure~\ref{fig:mu0}. In the function we have implemented for $\mu_R$ we used the central values of the parameters.

For the determination of $\sigma_{\rm R}$ and $\sigma_{\rm L}$ one has to consider that in general the convolution of a step function in $\mu$ with a Gaussian with standard deviation $\sigma$ will lead to a smoothed step function $1\pm{\rm erf}\left(\frac{x-\mu}{\sqrt{2}\sigma}\right)$.
Therefore, for $\sigma^2_{\rm L}$, we convolve Eq.~\eqref{eq:cassetta_norm_bis} with Eq.~\eqref{eq:gauss} for $n=1$ and obtain
\begin{align}
\sigma_L^2 &= \Sigma^2 + \sigma_{\rm ped}^2 = f^2R^2+\sigma_{\rm ped}^2\, ,
\end{align}
while for $\sigma^2_R$ we use Eq.~\eqref{eq:gauss} with $n=G_1$. Since Eq.~\eqref{eq:cassetta_norm_bis} falls off at $n\approx G_1$ with a slope $\sigma = \sqrt{G_1}$ deriving from a smoothed step function, by summing the variances, we obtain
\begin{align}
\sigma_R^2 &=  {f^2G_1+G_1f^2R^2+\sigma_{\rm ped}^2}.
\end{align}

\begin{figure}[h]
    \centering
    \includegraphics[width=0.7\columnwidth]{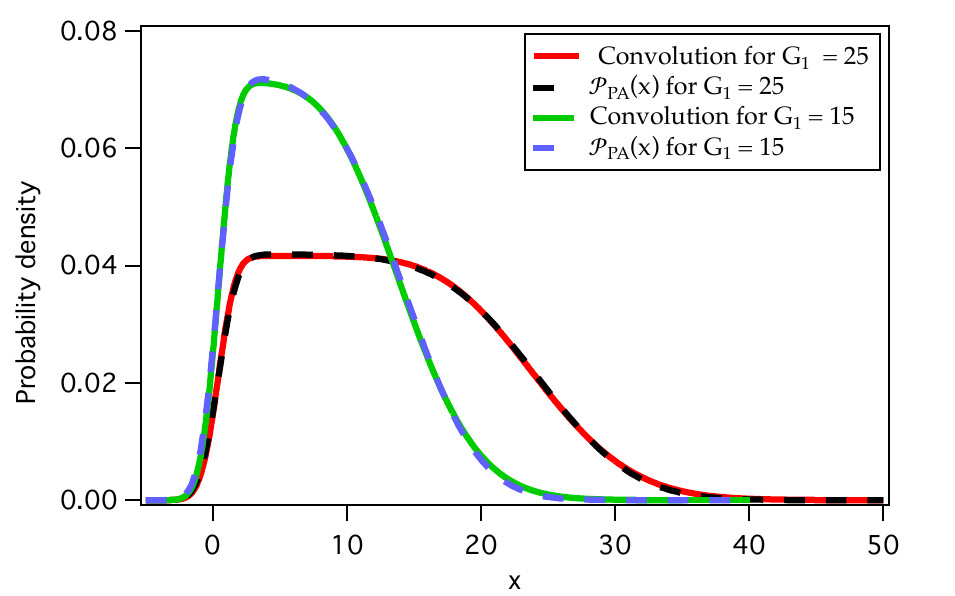}
    \caption{Comparison of the probability density function ${\mathscr P}_{\rm PA}(x)$ of Eq.~\eqref{eq:cassetta_conv_bis}, with the parameters defined in the text, with the direct convolution of the discrete probability in Eq.~\eqref{eq:cassetta_norm_bis} with Eq.~\eqref{eq:gauss} for two values of $G_1$. Here we have set $f = 1$, $R = 0.5$ and $\sigma_{\rm ped}/f = 1$.}
    \label{fig:cassetta_fit}
\end{figure}

A comparison of Eq.~\eqref{eq:cassetta_conv_bis}, with the parameters defined above, with a direct convolution of Eq. \eqref{eq:cassetta_norm_bis} with Eq. \eqref{eq:gauss} 
is shown in Figure \ref{fig:cassetta_fit} and demonstrates the validity of the functions and parameters obtained. 

\subsection{Poisson distribution for pre-pulses}
\label{sec:appendix_prepulse}
Pre-pulse events, due to photons passing through the photocathode and reaching directly the first dynode, are described in this model by a continuous scaled Poisson distribution.
The mean is therefore given by  $\langle{\mathscr P}_{\gamma \rm 1Dy}\rangle = G_2\cdot\langle G_3\rangle\cdot...\cdot \langle G_N\rangle$. Being $f =\langle G_2\rangle \cdot...\cdot \langle G_N\rangle$ as defined in Sec.~\ref{sec:succ_dynodes},
it follows that, in ADC units:
\begin{align} \langle{\mathscr P}_{\gamma \rm 1Dy}\rangle = f\frac{G_2}{\langle G_2\rangle} = f\zeta = f'.
\label{eq:gian_pre-pulse}
\end{align}
Because of this, also the variance $\sigma^2_{\gamma \rm 1Dy}$ is slightly modified to
\begin{align}
\sigma^2_{\gamma \rm 1Dy} = f'^2R'^2+\sigma^2_{\rm ped},
\end{align}
where (following the same procedure as in Section~\ref{sec:appendix1})
\begin{align}
\begin{split}
    R'^2 &= \frac{\sigma_2^2}{G_2^2} +\frac{\sigma_3^2}{ G_2\cdot \langle G_3\rangle^2}+...+\frac{\sigma_N^2}{ G_2\cdot \langle G_3\rangle\cdot ...\cdot \langle G_N\rangle^2}\\
    &=\frac{1}{\zeta}\left[R^2-\frac{\sigma_2^2}{\langle G_2\rangle^2}+\frac{\sigma_2^2}{\zeta\langle G_2\rangle^2}\right]\\&=\frac{R^2}{\zeta}-\frac{\left(\zeta-1\right)}{G_2} \, .
\end{split}
\label{eq:R'_bis}
\end{align}
For typical values of $\zeta \sim 1.1-1.3$ and $G_2 \sim 3-6$ the 
the second term on the right hand side in Eq.~\eqref{eq:R'_bis} represents a correction of the order of $10-15\%$ to the first term. Therefore we can make some approximations to express Eq.~\eqref{eq:R'_bis} without introducing the new parameter $G_2$. In particular, assuming that all the $N-1$ dynodes have the same gain $G_2$, and distinguishing between the gain $G_2$ at the second dynode and the average gain $\langle G_2\rangle$ at the following dynodes, the resolution due to the $N-1$ dynodes is~\cite{Wright}:
\begin{align}
    R'^2 \approx\frac{1}{G_2}\frac{\langle G_2\rangle}{\langle G_2\rangle-1}=\frac{1}{G_2-\zeta} .
\end{align}
Together with the approximation  $R'^2 \approx {R^2}/{\zeta}$ (Eq.~\eqref{eq:R'_bis}), one finds
\begin{align}
G_2\approx \zeta\left(\frac{1}{R^2}+1\right).
\label{eq:G_2}
\end{align}
Substituting Eq.~\eqref{eq:G_2} in Eq.~\eqref{eq:R'_bis} defines $R'^2$ using only the previously defined parameters $R$~and~$\zeta$:
\begin{align}
    R'^2 \approx \frac{R^2}{\zeta}\left[1- \frac{\zeta-1}{R^2+1}\right].
\end{align}

\bibliographystyle{spphys}
\bibliography{main}

\end{document}